\documentclass[preprint,3p]{elsarticle}

\usepackage{lineno,hyperref}
\modulolinenumbers[5]
\usepackage{xcolor}
\usepackage{booktabs}
\usepackage{arydshln}
\usepackage{multirow}
\usepackage{amsmath,amssymb}
\usepackage{float}
\usepackage{subfig}
\usepackage{graphics}
\usepackage{graphicx}
\usepackage{comment}
\usepackage{smartdiagram}
\journal{Nuclear Physics B}

\begin{document}

\begin{frontmatter}

\title{Muon $g-2$: A review}

\author[manchester]{Alex Keshavarzi}
\ead{alexander.keshavarzi@manchester.ac.uk}

\author[sjtu]{Kim Siang Khaw}
\ead{kimsiang84@sjtu.edu.cn}

\author[kyushu]{Tamaki Yoshioka}
\ead{yoshioka@phys.kyushu-u.ac.jp}

\address[manchester]{Department of Physics and Astronomy, The University of Manchester, Manchester M13 9PL, United Kingdom}
\address[sjtu]{Tsung-Dao Lee Institute and School of Physics and Astronomy, Shanghai Jiao Tong University, Shanghai 200240, China}
\address[kyushu]{Research Center for Advanced Particle Physics, Kyushu University, Fukuoka 819-0395, Japan}

\begin{abstract}
The muon magnetic anomaly, $a_{\mu }=(g_{\mu }-2)/2$, plays a special role
in the history of the Standard Model (SM). Precise calculations and measurements
of this fundamental quantity provide a stringent test of the SM and a window
to the physics beyond. In light of the first result published by the Fermilab
Muon $g-2$ experiment, this article reviews the current status of the experimental
measurement and theoretical prediction of the muon anomalous magnetic moment.
It also provides an overview on resulting constraints of associated physics
beyond the SM (BSM), future muon $g-2$ experiments and on the progress
of experiments that are closely connected to the physics of the muon
$g-2$, such as measurements that could provide complementary information
about BSM physics in the muon sector.
\end{abstract}

\begin{keyword}
muon, anomalous magnetic moment, precision measurement, beyond the standard model
\end{keyword}

\end{frontmatter}



\section{Introduction}

The magnetic moment of the muon plays a special role in the establishment of the Standard Model (SM) of elementary particles. For an elementary particle with intrinsic angular momentum (spin, $\vec{S}$) and charge $q$, its magnetic moment $\vec{\mu}$ is given (in natural units) by
\begin{equation}
    \vec{\mu} = g\frac{q}{2m}\vec{S} \, ,
\end{equation}
where $g$ is the gyromagnetic ratio and $m$ is the mass of the particle. From his original formulation of quantum mechanics, Dirac predicted $g=2$ for the electron (and, consequently, any spin-$\frac{1}{2}$ elementary particle) in 1928~\cite{Dirac:1928ej}. However, in the description of relativistic quantum field theories, this quantity receives contributions from radiative corrections, where the interaction of the elementary particle with a photon is modified by additional interactions with virtual particles. These quantum fluctuations modify $g$, where interactions with virtual particles increase its value from the tree-level prediction of $g=2$. The first order correction from quantum electrodynamics (QED) was performed by J.~Schwinger in 1948 and was found to be an increase to $g$ of exactly $\alpha/\pi$~\cite{Schwinger:1948iu}, where $\alpha$ is the fine-structure constant. Experimental confirmation of this followed from Kusch and Foley in the same year~\cite{Kusch:1948mvb}. For the charged leptons ($l=e,\mu,\tau$), the magnetic anomaly $a_l$ is defined as the fractional deviation from Dirac's prediction of $g_l=2$: 
\begin{equation}
a_{l} = (g_{l}-2)/2\, .
\end{equation}
This quantity continues to serve as a long-standing test of the SM to this day. Comparisons with experimental (EXP) measurements $a_l^{\rm EXP}$ result in studies of the magnetic moments of leptons being a powerful indirect search of new physics. The SM calculations, $a_l^{\rm SM}$, require determining the modifications to $g$ from virtual SM particles up to a sufficient order. Should precision calculations of $a_l^{\rm SM}$ and precision measurements of $a_l^{\rm EXP}$ disagree, it could serve as an indication of beyond the SM (BSM) contributions, where BSM virtual particles or forces may be providing additional contributions to $a_l$. 

Continued studies of the electron anomaly, $a_e$, have resulted in both its theoretical prediction and experimental measurement being two of the most precisely measured quantities in particle physics. The most recent experimental measurement of $a_e^{\rm EXP} = 1\,159\,652\,180.73(28) \times 10^{-12}$~\cite{Hanneke:2008tm} is precise to 0.24 parts-per-billion (ppb). The SM prediction $a_e^{\rm}$, whilst being robust in its own calculation~\cite{Aoyama:2012wj,Aoyama:2017uqe}, is sensitive to the experimentally measured value of $\alpha$. Over roughly the last decade, three separate measurements of $\alpha$ have been performed, two via rubidium (Rb) atomic
interferometry~\cite{Bouchendira:2010es,Morel:2020dww} and one using caesium (Cs) atomic interferometry~\cite{Parker:2018vye}. The results of these measurements for both $\alpha$, $a_e^{\rm SM}$ and the difference $\Delta a_e = a_e^{\rm EXP} - a_e^{\rm SM}$ are given in Table~\ref{tab:ae}. These values vary and result in differing values of $a_e^{\rm SM}$ and $\Delta a_e$ that imply both negative and positive tensions with $a_e^{\rm EXP}$ depending upon the measured value of $\alpha$. These results have invoked much theoretical work into explaining the deviations in the electron sector. A negative $\Delta a_{e}$ could, for example, require BSM models that break lepton flavor universality.
\begin{table}
\centering
\begin{tabular}{ccccc}
\hline
Measurement & $\alpha^{-1}$ & $a_e^{\rm SM} \times 10^{12}$ & $\delta a_e^{\rm SM}/a_e^{\rm SM}$ (ppb) & $\Delta a_e  \times 10^{13} $ \\
\midrule
$\alpha_{\rm Rb}$~(2010)~\cite{Bouchendira:2010es} & $137.035999037(91)$ & $1\,159\,652\,182.032(720)$ & 0.62 & $-13.0\pm
7.7\,(-1.7\sigma)$ \\
$\alpha_{\rm Cs}$~(2018)~\cite{Parker:2018vye} & $137.035999046(27)$ & $1\,159\,652\,181.61(23)$ & 0.19 & $-8.8 \pm
3.6\,(-2.5\sigma)$\\
$\alpha_{\rm Rb}$~(2020)~\cite{Morel:2020dww} & $137.035999206(11)$ & $1\,159\,652\,180.252(95)$ & $0.08$ & $ 4.8\pm3.0\, (+1.6\sigma)$ \\
\hline
\end{tabular}
\caption{Results for $a_e^{\rm SM}$ and $\Delta a_e$ resulting from different measurements of $\alpha$.} 
  \label{tab:ae}
\end{table}

This article reviews the current status of the experimental measurements and theoretical calculations of the muon magnetic anomaly $a_\mu$.\footnote{The current status of $a_\tau$ is insignificant due to the insufficient accuracy of $a_{\tau}^{\rm EXP}$ and is therefore not discussed here (see~\cite{Keshavarzi:2019abf} for more details).} The larger mass of the muon induces several interesting features compared to the electron. Mass scaling arguments mean that the muon is typically $(m_\mu/m_e)^2\sim 4\times10^4$ times more sensitive to BSM effects from heavy new particles than the electron. The larger mass of the muon also enhances the hadronic sector contribution to $a_\mu$ relative to $a_e$, resulting in a larger relative uncertainty on $a_{\mu}^{\rm SM}$ due to the non-perturbative nature of low energy strong interactions. This larger uncertainty makes $a_\mu$ less sensitive to $\alpha$ and, therefore, unaffected by the current inconsistencies between its experimental measurements. Coupled together, these make $a_\mu$ a suitable and interesting candidate to probe the SM and search for BSM physics. Following a review of the experimental efforts and most recent results, current status of $a_{\mu}^{\rm SM}$ will be discussed. An overview on the possible sources of BSM contributions and resulting constraints on several BSM scenarios are also given. Future plans of the Fermilab and J-PARC experiments will additionally be discussed, together with experiments that are related and could provide complementary information to muon $g-2$ physics.

\section{Experimental measurements of muon $g-2$}

\subsection{Historical development}

The first-ever measurement of the muon magnetic moment was performed by the Columbia Nevis group in the 1950s as an investigation of the parity violation of weak muon decay and resulted in a measurement of $g$ that was accurate to 5\%~\cite{Garwin:1957hc}. In the same year, a measurement at the University of Liverpool improved the precision of $g$ to less than 1\%~\cite{Cassels:1957}. An improved measurement at the Columbia Nevis Lab measurement in 1960 achieved a precision of 6.5\% on $a_{\mu}$~\cite{Garwin:1960zz}. 

Starting from 1959, Lederman and collaborators launched a muon $g-2$ measurement campaign at CERN. The first of a series of three CERN experiments, CERN I, yielded a 0.4\% measurement of $a_{\mu}$~\cite{Charpak:1962zz, Charpak:1965zz}. Realizing the need to use relativistic muons to increase the observation time of the spin precession for better precision, storage ring techniques were pioneered for the second experiment at CERN II~\cite{Combley:1974tw}. To eliminate the motional magnetic field effect due to the use of focusing electric quadrupoles, CERN III used the ``magic momentum" approach, where muons with 3.1~GeV/c momentum are stored for the measurement of $a_{\mu}$ resulting in a cancellation of electric field contributions to first order~\cite{Bailey:1978mn}. Combining all the measurements taken at CERN, an impressive 7.3~ppm precision on $a_{\mu}$ was achieved~\cite{Bailey:1978mn}. 

Following the end of the CERN campaign, the Brookhaven National Laboratory (BNL) employed a dedicated superferric superconducting storage ring magnet with a more uniform magnetic field, a muon injection approach utilizing a passive inflector magnetic and pulsed magnetic kickers. These (and other changes with respect to previous measurements) resulted in significant improvements over the predecessor measurements. After the data taking concluded in 2001, the BNL muon $g-2$ collaboration delivered a result for $a_\mu$ that was precise to 0.54~ppm~\cite{Bennett:2006fi}. A summary of the evolution of the experimental techniques is given in Table~\ref{tab:cern_and_bnl}.

\begin{table}[!t]
\begin{tabular}{cccccc} \toprule 
 Experiment & Magnet & Approach & $\gamma_{\mu}$ & $\delta a_{\mu}/a_{\mu}$ & Ref. \\ \hline 
 CERN I (1965) & Long dipole magnet, $B=1.6$~T & $\mu$ injection & 1 & 0.4\% & \cite{Charpak:1962zz,Charpak:1965zz}\\   
 CERN II (1974) & $R=2.5$~m  storage ring, $B=1.71$~T & $p$ injection & 12 & 270~ppm &  \cite{Combley:1974tw}  \\  
 CERN III (1978) & $R=7.1$~m storage ring, $B=1.47$~T & $\pi$ injection & 29.3 & 7.3 ~ppm & \cite{Bailey:1978mn}   \\  
 BNL (2006) & $R=7.1$~m storage ring, $B=1.45$~T & $\mu$ injection & 29.3 & 0.54~ppm & \cite{Bennett:2006fi}\\ 
 FNAL Run-1 (2021) & $R=7.1$~m storage ring, $B=1.45$~T & $\mu$ injection & 29.3 & 0.46~ppm & \cite{Abi:2021gix}\\  \bottomrule
\end{tabular}
\caption{A summary of the development of the muon $g-2$ experiments.}
\label{tab:cern_and_bnl}
\end{table}

\subsection{Principles of a storage ring muon $g-2$ experiment}

Modern muon $g-2$ experiments inject spin-polarized muons into a magnetic storage ring and track the spin evolution as muons circulate. The determination of $a_{\mu}$ is achieved by measuring the anomalous precession frequency $\omega_a$, defined as the difference between the spin precession frequency $\omega_s$ relative to the cyclotron frequency $\omega_c$ for muons orbiting a highly uniform magnetic field $\vec{B}$.
Expressed using above-mentioned quantities, in the absence of an electric field and in the limit of planar muon orbits $\vec{\beta}\cdot\vec{B}\approx0$,
\begin{equation}
\vec{\omega}_a  \equiv \vec{\omega}_s - \vec{\omega}_c = -a_{\mu} \frac{q\vec{B}}{m_\mu}\, ,
\label{eq:omegaa}
\end{equation}
where $\vec{\beta}$ denotes the muon velocity. It follows that a precise determination of $a_{\mu}$ requires high-precision knowledge of the $\omega_a$ frequency and magnetic field $\vec{B}$. 

The $\omega_a$ frequency is imprinted in the modulation of the decay-positron energy spectrum as muons circulate in the ring. Due to the parity-violating weak decay $\mu^+ \rightarrow e^+{\bar{\nu}}_{\mu}\nu_e$, in the muon rest frame, high energy positrons are emitted preferentially in the direction of the muon spin. In the laboratory frame, the energy spectrum depends on the relative angle between the muon spin and momentum: when the muon spin and momentum are aligned, the energy spectrum is the hardest, and vice versa.
Therefore, in the absence of effects from the spatial and temporal motion of the beam, the number $N(t)$ of higher-energy positrons observed by detectors placed at the inner part of the storage ring is given by
\begin{equation}\label{eq:5par}
N(t) = N_{0}e^{-t/\tau}[1 + A \cos (\omega_a t + \phi)]\, ,
\end{equation}
where $N_{0}$ is a normalization, $\tau$ is the time-dilated muon lifetime, $A$ is the muon decay asymmetry, and $\phi$ is the initial $g-2$ phase. 

In the presence of an electric $\vec{E}$ from focusing electric quadrupoles and non-planar muon orbits, equation~\eqref{eq:omegaa} is modified as relativistic particles feel a motional magnetic field proportional to $\vec{\beta}\times\vec{E}$:
\begin{equation}
\vec{\omega}_{a}\equiv\vec{\omega}_{s}-\vec{\omega}_{c} =-\frac{q}{m_{\mu}}\Bigg[ a_{\mu} \vec{B}-a_{\mu}\left(\frac{\gamma}{\gamma+1}\right)(\vec{\beta} \cdot \vec{B}) \vec{\beta} \left.-\left(a_{\mu}-\frac{1}{\gamma^{2}-1}\right) \frac{\vec{\beta} \times \vec{E}}{c} \right].
\label{eq:omega_aFull}
\end{equation}
For horizontally circulating muons in a vertical magnetic field, $\vec{\beta}\cdot\vec{B} = 0$ and, in the ``magic momentum" approach where the muons are chosen to have a central momentum of 3.1~GeV/c momentum, it follows that $\gamma_\mu = \sqrt{(1+1/a_\mu)} \approx 29.3$ additionally reduces the third term to zero. However, there persists a small momentum spread of remaining muons away from the magic momentum and a small amount of vertical pitching, corresponding to the corrections arising from second and third terms of equation~\eqref{eq:omega_aFull} respectively. Due to these and other effects that arise as a result of the storage environment, the function in equation~\eqref{eq:5par} is not sufficient to describe the corresponding motion of the stored beam, as detectors are sensitive to such effects. As such, extra terms must be necessarily added to equation~\eqref{eq:5par} to describe the relevant adjustments. The magnitudes of these can be determined precisely from analysis and corrected for, resulting in corresponding uncertainties to be applied to the measured $\omega_a$.  

The magnetic field $|\vec{B}|$ is measured using nuclear magnetic resonance (NMR) techniques where NMR probes are placed around the storage ring to measure and monitor the magnetic field during the experiment. The field is expressed in terms of the proton Larmor precession frequency $\omega_{p}$ as
\begin{equation}
    \hbar \omega_{p} = 2\mu_{p}B \, ,
    \label{eq:omegap}
\end{equation}
where $\mu_{p}$ is the proton magnetic moment.
Combining equation~\eqref{eq:omegaa} and equation~\eqref{eq:omegap} and using the definition of electron magnetic moment, $\mu_{e}=g_{e}e\hbar/4m_{e}$, one obtains
\begin{equation} \label{eq:amu}
    a_{\mu} = \frac{\omega_a}{\omega_p} \frac{\mu_{p}}{\mu_{e}} \frac{m_{\mu}}{m_e} \frac{g_{e}}{2} \, ,
\end{equation}
where $\mu_{p}/\mu_{e}$ is the proton-to-electron magnetic moment ratio and $m_{\mu}/m_{e}$ is the muon-to-electron mass ratio. The last three ratios in equation~\eqref{eq:amu} are determined from independent experiments and the ratio $\omega_a/\omega_p$ is the one provided by a muon $g-2$ experiment. Historically, the expression
\begin{equation}
    a_{\mu} = \frac{R}{\lambda-R}
\end{equation}
has also been utilized~\cite{Bailey:1978mn,Bennett:2006fi}, where $R=\omega_a/\omega_p$ and $\lambda=\mu_{\mu}/\mu_{p}$ is the muon-to-proton magnetic moment ratio. The current status of measurements of these fundamental constants is reported in Sec.~\ref{sec:relatedexps}.

\subsection{Fermilab experiment}

\begin{figure}[htbp]
\centering
    \includegraphics[width=0.85\textwidth]{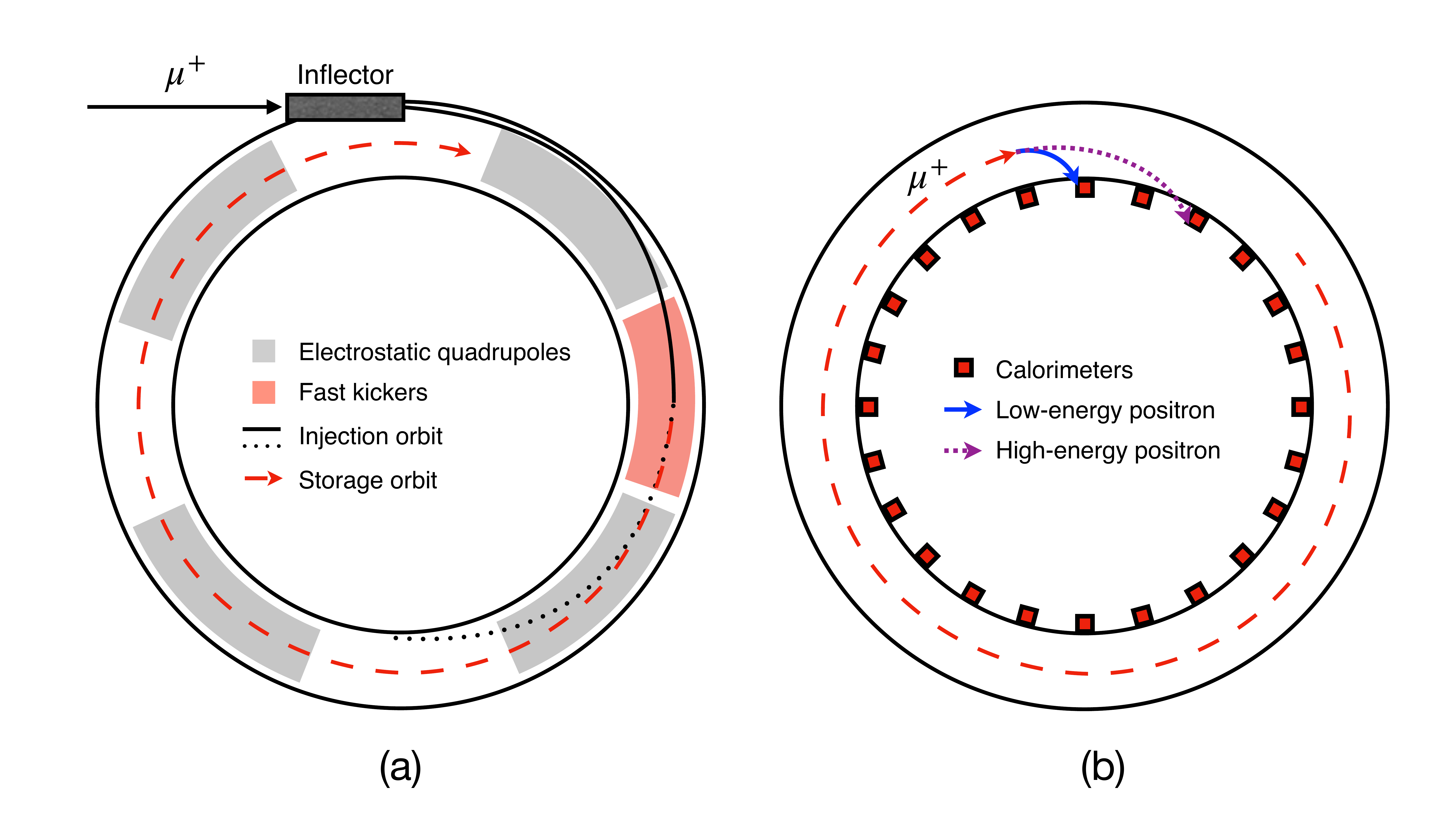}
  \caption{\small Sketches of the (a) muon beam injection and (b) positron detection system in the Fermilab Muon $g-2$ experiment. For the beam injection, the muon first traverses through a superconducting inflector and is deflected by a series of fast kickers. The horizontal confinement is provided by the 1.45-T vertical magnetic field of the ring while the vertical confinement is provided by the electrostatic quadrupoles covering 43\% of the ring. For the positron detection, 24 calorimeters are installed in the inner part of the ring. Depicted are two decay-positron trajectories with different energies from muons decaying at the same location.} \label{fig:fermilabg2_overview}
\end{figure}

The Fermilab Muon $g-2$ experiment~\cite{Grange:2015fou} employs the same approach as at BNL~\cite{Bennett:2006fi}, but with much-improved instrumentation for the magnetic field and muon spin precession frequency measurements. The 7.112~m radius, 1.45~T superconducting storage ring at BNL~\cite{Danby:2001eh} was relocated to Fermilab. At Fermilab, intense bunches of polarized positive muons having a central momentum of 3.1~GeV/c are injected into the muon storage ring at an average rate of 11.4~Hz. This beam has a negligible hadron contamination, but a non-negligible fraction of positrons. After traversing through a narrow channel of superconducting inflector magnet and one-quarter of the storage ring, the beam is deflected by a fast pulsed-kicker magnet~\cite{Schreckenberger:2021kur} onto the intended storage orbit. There are four electrostatic quadrupole (ESQ) sections installed symmetrically around the ring to confine the beam vertically. Only a few percent of the muons remain after a few complete orbits of the storage ring (the cyclotron period is 149.2~ns). The unstored muons, together with the positron beam contamination, generate a prompt background in the calorimeter systems that is anticipated in the detector and electronics designs. A schematic of the beam injection and storage of the experiment is given in Figure~\ref{fig:fermilabg2_overview}(a).

The measurement of the $\omega_a$ is performed by 24 calorimeter stations positioned equidistantly around the inner radius of the storage ring as depicted in Figure~\ref{fig:fermilabg2_overview}(b). Each calorimeter is located in a scalloped region of the vacuum chamber system. These calorimeters are based on the use of Cherenkov radiator of PbF$_2$ crystals from which signals are read out via silicon photo-multipliers (SiPM)~\cite{Fienberg:2014kka,Kaspar:2016ofv,Khaw:2019yzq}. Each consisting of a $9\times6$ array of PbF$_2$ crystals, these fast and segmented calorimeters provide an excellent timing resolution (sub-ns) and spatial resolution for distinguishing individual positron events. A new laser-based gain monitoring system~\cite{Muong-2:2019hxt} is also employed to stabilize the calorimeter energy measurement and to enable better detector gain corrections. An in-vacuum tracking system based on straw trackers~\cite{Stuttard:2017fti, Mott:2017aon,King:2021hst} is installed at two locations of the storage ring, each upstream of a calorimeter. Details about $\omega_a$ measurement are given in~\cite{Albahri:2021ixb}.

These are significant upgrades compared to the BNL experiment where the calorimeters were read out as a single monolithic block, the calibration measurements based on nitrogen laser gain monitoring system were not being used due to larger-than-expected data fluctuations, the traceback wire chambers were installed outside of the vacuum chamber resulting in a larger position reconstruction uncertainty due to material and multiple scattering effect.

The storage ring magnetic field is measured by means of the nuclear magnetic resonance (NMR) technique where 378 fixed NMR probes are placed in 72 azimuthal locations around the storage ring to monitor the field around the muon storage region, and 17 NMR probes are installed in an in-vacuum trolley to map the field in the storage region. A precision calibration probe consisting of a high-purity cylindrical water sample is used for absolute calibration~\cite{Flay:2021jrs}. While the design of the trolley and fixed NMR probes largely follow the design from the BNL experiment, trolley electronics, trolley position encoders, and controllers are upgraded for the Fermilab experiment~\cite{Corrodi:2020sav}. The RMS of the magnetic field variation around the storage ring was about 10~ppm after passive shimming, approximately 3 times better than that of a typical magnetic field scan at BNL. Details about the field measurements are elaborated in~\cite{Albahri:2021kmg}. 

Another important feature of the Fermilab measurement compared to the BNL experiment is the extensive use of three state-of-the-art and complementary particle tracking simulation programs to understand and estimate various systematic uncertainties. These are the Geant4 simulation toolkit~\cite{GEANT4:2002zbu} and accelerator simulation toolkits BMAD~\cite{Sagan:2006sy} and COSY~\cite{Makino:2006sx}. Details about simulations and estimation of systematics based on these programs are outlined in~\cite{Albahri:2021mtf}.

\begin{figure}[htbp]
\centering
    \includegraphics[width=0.7\textwidth]{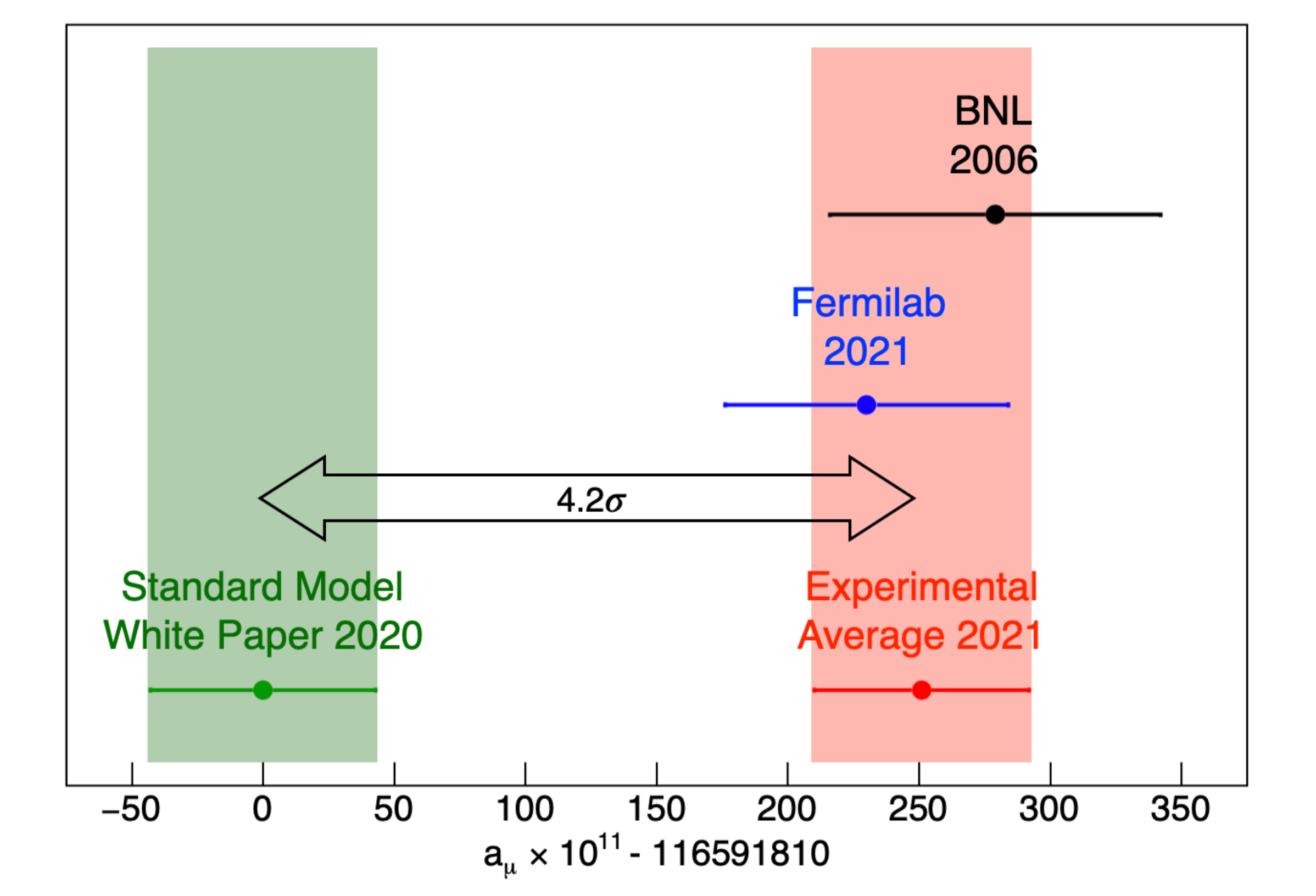}
  \caption{\small Comparison between experimental values and theoretical prediction of $a_{\mu}$.}\label{fig:latest_g-2} 
\end{figure}

The Fermilab experiment was fully commissioned in Winter 2018 and the first physics run was completed in late June 2018. Analyses of the Run-1 data proceeded in a blind fashion and were unblinded in early 2021. Details of the analyses are discussed in~\cite{Abi:2021gix,Albahri:2021mtf,Albahri:2021ixb,Albahri:2021kmg} and provided here are key elements of the analyses. Determination of the muon anomalous precession frequency was performed by 6 independent teams using sophisticated analysis procedures~\cite{Albahri:2021ixb}. In the published result, two independent positron reconstruction algorithms were used in four independent measurements of $\omega_{a}$. To obtain optimal statistical sensitivity, each positron event was weighted by its energy-dependent decay asymmetry. Three data-driven techniques were developed for removing pileup positron events in the raw time and energy distribution. Short-term and long-term gain fluctuations were corrected for using laser-based calibration system. The major systematics on $\omega_{a}$ arise from uncertainty in the pileup correction, gain corrections, modelling of the functional form of the beam dynamics model in the fit to the positron time distribution. The total systematic uncertainty for the fitted $\omega_{a}$ is 56~ppb for Run-1.

The measured value of $\omega_{a}$ described above requires 4 additional corrections before it can be interpreted as the $\omega_{a}$ in equation~\eqref{eq:amu}. These corrections are the electric-field correction $C_{e}$ (arising from the last term in equation~\eqref{eq:omega_aFull} due to non-magic momentum muons), the pitch correction $C_{p}$ (arising from the vertical betatron oscillations that lead to a nonzero
average value of $\vec{\beta}\cdot\vec{B}$), the lost muon phase correction $C_{ml}$ (arising from any bias in the average muon $g-2$ phase of lost muons compared to stored muons during the measurement period), and the phase-acceptance correction $C_{pa}$ (arising from the slow change in the muon beam distribution coupled to the decay coordinate dependent muon $g-2$ phase). The last two corrections are expected to be smaller in Run-2 as damaged ESQ resistors in Run-1, which caused a significant increase in the lost muon rate and the instability in the beam distribution, were replaced. The total systematic uncertainty for corrections to $\omega_{a}$ is 93~ppb for Run-1 datasets.

Determination of the magnetic field strength in the storage region was done by performing a sequence of measurements using high-precision magnetometers. First, each of the 17 NMR probes of the in-vacuum trolley~\cite{Corrodi:2020sav} was calibrated with a plunging probe. Then the magnetic field in the storage region was mapped approximately every 3 days using the trolley probes. The 378 fixed NMR probes located around the ring were synchronized to each trolley measurement. The roles of these probes are to provide a feedback to the magnet power supply to stabilize the field and to track the evolution of the field between the trolley measurements. To obtain the magnetic field experienced by the muon ensemble, the measured field maps were weighted by the spatial and temporal distributions of the muons included in the $\omega_{a}$ analysis. Finally, corrections were made for transient fields generated by the storage ring components ($B_{k}$ for eddy currents in the kicker plates and $B_{q}$ for mechanical vibration of the charged ESQ plates) that are not resolved by the field tracking and not present during the trolley measurement. The total systematic uncertainty for the muon distribution averaged magnetic field is 56~ppb and for the field correction is 99~ppb for Run-1. A summary of the uncertainties in Run-1 data analysis is provided in Table~\ref{tab:uncertainty_summary}. 

\begin{table}[htbp]
\centering
\small
\begin{tabular}{ccccc} \toprule 
 Quantity & Correction (ppb) & Uncertainty (ppb) & Fermilab goal (ppb) & J-PARC goal (ppb)\\ \hline 
 $\omega_a$ (statistical) & - & 434 & 100 & 450\\  \hline 
 $\omega_a$ (systematic) & - & 56 & 70 & $<40$\\ \hline 
 $C_{e}$ & 489 & 53 & - & - \\
 $C_{p}$ & 180 & 13 & - & - \\
 $C_{ml}$ & -11 & 5 &  - & - \\
 $C_{pa}$ & -158 & 75 & - & -\\ \hline 
 $\omega_p$ (systematic) & - & 56 & 70 & 56\\ \hline 
 $B_{k}$ & -27 & 37 & - & -\\
 $B_{q}$ & -17 & 92 & - & -\\ \hline 
 Total systematic & - & 157 & 100 & $<70$\\ \hline 
 Fundamental constants & - & 25 & - & -\\ \hline 
 Total & 544 & 462 & 140 & $<460$ \\
   \bottomrule
\end{tabular}
\caption{Values and uncertainties of the Run-1 data analysis compared to the Fermilab uncertainty goal~\cite{Grange:2015fou}}
\label{tab:uncertainty_summary}
\end{table}

The Run-1 result from the Fermilab Muon $g-2$ collaboration~\cite{Abi:2021gix,Albahri:2021mtf,Albahri:2021ixb,Albahri:2021kmg} gives
\begin{equation}
a_{\mu}^{\rm{FNAL}}= 116~592~040(54) \times 10^{-11}~(0.46\,\rm{ppm})~.
\end{equation}
It exhibits a 3.3$\sigma$ tension with the value published by the Theory Initiative~\cite{Aoyama:2020ynm} that will be reviewed in Section~\ref{sec:g2theory}. After a statistically consistent combination with the final BNL result~\cite{Bennett:2006fi}, the new experimental average is
\begin{equation}
  a_{\mu}^{\rm{EXP}}= 116~592~061(41) \times 10^{-11}~(0.35\,\rm{ppm})~, 
\end{equation}
and the tension is increased to 4.2$\sigma$, as shown in Figure~\ref{fig:latest_g-2}. A new analysis of the data from Run-2 and Run-3 (amounting to $\sim$ 3 times the Run-1 data) is expected to be completed within the next few years. The uncertainty is expected to be roughly half of the Run-1, based on the following improvements achieved after Run-1:
\begin{itemize}
    \item \textbf{Storage ring conditions}: While Run-1 has 4 different storage ring conditions (electric quadrupole and kicker HV settings), Run-2 and Run-3 each have only a single setting for the quadrupole HV setting. Moreover, the kicker system was improved to withstand higher voltages for the second half of Run-3 and, as a result, the stored beam is more centered compared to Run-1. It was also discovered after Run-1 that two of the resistors connected to the electrostatic quadrupoles were faulty, consequently affecting the stability of the necessary beam focusing. Corresponding effects from these faulty resistors are comprehensively dealt with in the analysis of the Run-1 data. These resistors were replaced and the beam stability for all following measurement periods was re-established.
    \item \textbf{Stability and gradient of storage ring magnet temperature}: After Run-1, a new set of NMR probes at optimized locations were selected, improving the stability of the average magnetic field as a function of time. Additionally, a larger-than-design heat load from various sub-systems prompted a better temperature stabilization system of the experimental hall and insulation of the superconducting magnet was installed during Run-2 to further reduce the temperature gradient around the azimuth of the storage ring. During Run-3, a more stable experimental hall and superconducting magnet temperature was achieved. Therefore, there was less variation in the magnetic field throughout Run-3 and the systematics related to the field tracking is reduced.
\end{itemize}
The Fermilab Muon $g-2$ experiment has also completed the Run-4 data-taking campaign and the Run-5 data-taking has began in late 2021. There is also a possibility in performing the $a_{\mu}$ measurement using negative muons after 2022. This will require polarity switching of the beamline components and the storage ring components. 

Besides the magnetic anomaly measurement, other physics studies of interest at the Fermilab experiment include tests of Lorentz-symmetry and CPT-symmetry~\cite{Gomes:2014kaa, Quinn:2019ppv}. The CPT-symmetry test can be performed by comparing $a_{\mu}$ from $\mu^{+}$ and $\mu^{-}$ measurements, while Lorentz symmetry can be tested by searching for a sidereal variation in the anomalous precession frequency of the muon.

\section{Theoretical calculations of $a_{\mu}$}\label{sec:g2theory}

\begin{figure}[!t]
\centering
    \includegraphics[width= 0.7\textwidth]{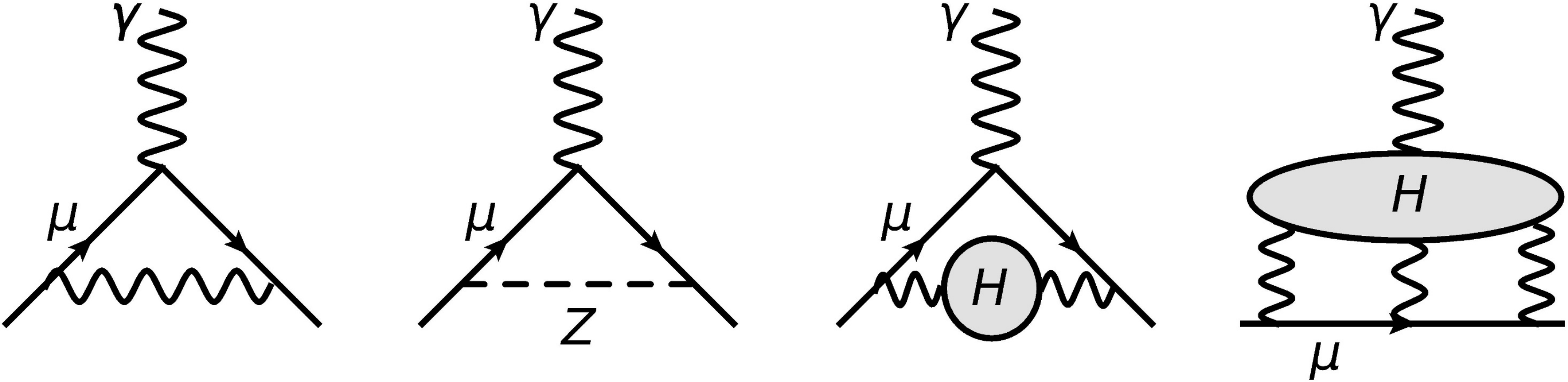}
  \caption{\small  Feynman diagrams of SM contributions to $a_\mu$. The diagrams shown (from left to right) are the one-loop QED diagram, the one-loop EW process involving $Z$-boson exchange, the leading-order HVP diagram and HLbL contributions.}\label{fig:SMdiagrams} 
\end{figure}

In this section, the SM value of the anomalous magnetic moment against which the experimental value~\cite{Abi:2021gix} is compared to will be reviewed. The SM prediction of the anomalous magnetic moment is determined from the sum of all sectors of the SM:
\begin{equation} \label{eq:alSMeq}
a_{\mu}^{\rm SM} = a_{\mu}^{\rm QED} + a_{\mu}^{\rm EW} + a_{\mu}^{\rm HVP} + a_{\mu}^{\rm HLbL}  \, ,
\end{equation}
where $a_{\mu}^{\rm QED}$ are the QED contributions, $a_{\mu}^{\rm EW}$ are the electro-weak (EW) contributions, $a_{\mu}^{\rm HVP}$ are the hadronic vacuum polarization (HVP) contributions and $a_{\mu}^{\rm HLbL}$ are due to contributions from hadronic light-by-light (HLbL) scattering. Examples of such processes are shown in Figure~\ref{fig:SMdiagrams}. The uncertainty of $a_{\mu}^{\rm SM}$ is completely dominated by the hadronic contributions due to the non-perturbative nature of the low energy strong interaction. In recent years, $a_{\mu}^{\rm SM}$ has been thoroughly scrutinized and reevaluated by The Muon $g-2$ Theory Initiative~\cite{Aoyama:2020ynm}, an international collaboration determined on providing a community-approved consensus for the value of the theoretical prediction with an improved overall precision. 

\subsection{The QED contributions}

\begin{figure}[!t]
\centering
    \includegraphics[width= 0.8\textwidth]{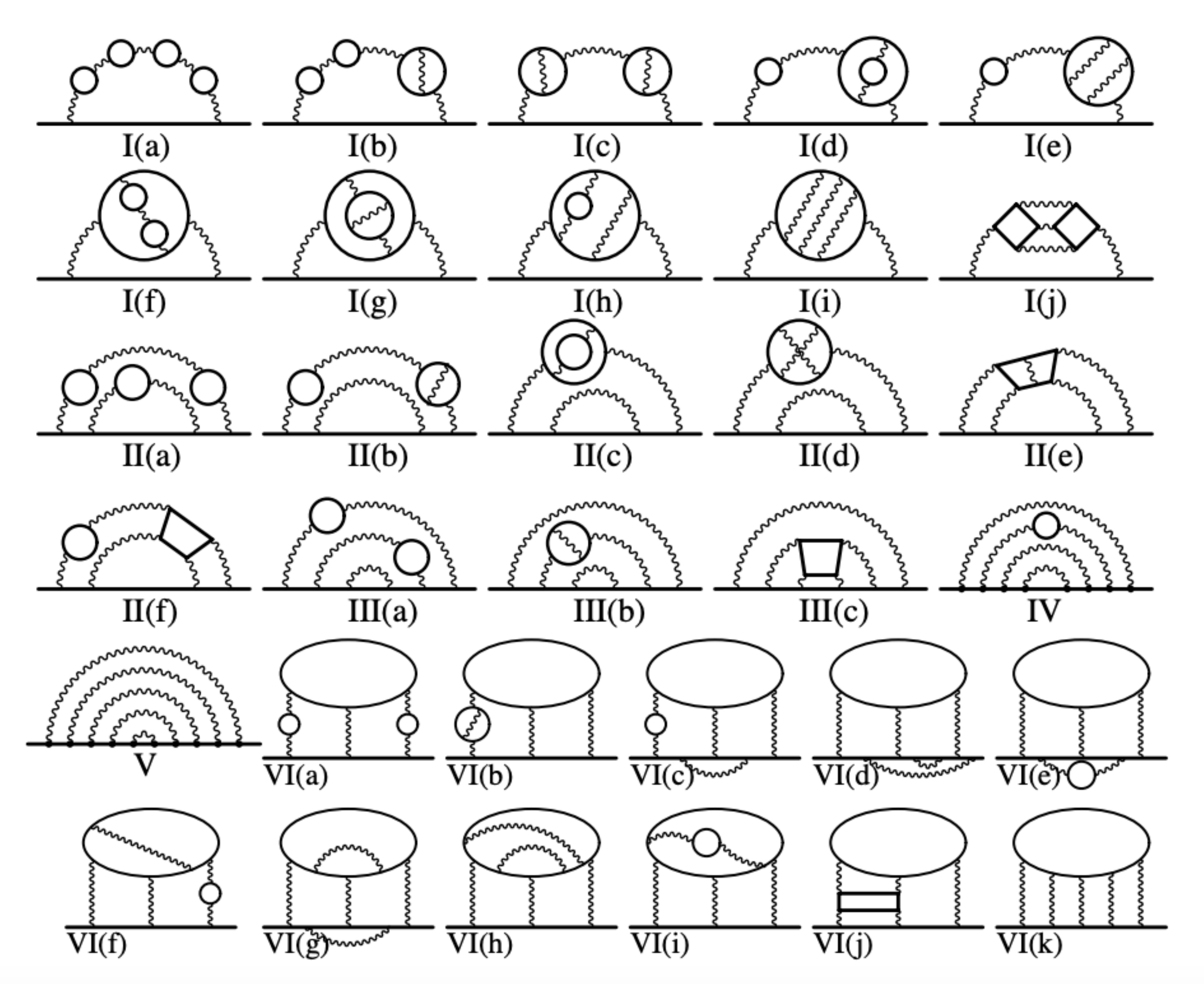}
  \caption{\small  Five-loop QED diagrams. There are, in total, 12,672 diagrams contributing to $a_{\mu}^{\rm QED}$. The straight and wavy lines represent lepton and photon propagators,
respectively. This figure is reprinted from (and further described in)~\cite{Aoyama:2012wk}.}\label{fig:5loopQED} 
\end{figure}
The QED contributions to $a_\mu$ include all contributions from leptons and photons alone and have been fully calculated up to five-loop order.
All contributions up to and including four-loop have been determined and verified by different groups, from both numerical and analytical calculations (see~\cite{Aoyama:2020ynm} for more details). The four-loop universal contribution has been impressively calculated analytically up to a precision of 1100 digits~\cite{Laporta:2017okg} and is consistent with the numerical determination~\cite{Aoyama:2014sxa}. The entire five-loop contribution, totaling 12,672 Feynman diagrams which are shown in Figure~\ref{fig:5loopQED}, has been fully calculated numerically~\cite{Aoyama:2012wk,Aoyama:2019ryr} with independent cross-checks~\cite{Volkov:2019phy,Kataev:1991cp,Laporta:1994md,Baikov:2013ula}. The value for the QED contributions is found to be
\begin{equation} \label{eq:alQED}
a_{\mu}^{\rm QED} =  116~584~718.931(104) \times 10^{-11} \, ,
\end{equation}
where the given error is the quadrature sum of uncertainties due to the $\tau$-lepton mass, four-loop QED, five-loop QED, an estimate of the six-loop QED~\cite{Aoyama:2020ynm,Aoyama:2012wk,Aoyama:2019ryr} and the fine-structure constant $\alpha$~\cite{Parker_2018}.\footnote{This value for $a_{\mu}^{\rm QED}$ is obtained using the measurement of $\alpha$ from caesium interferometry~\cite{Parker_2018}. With the uncertainty of $a_{\mu}^{\rm QED}$ dominated by the six-loop estimate~\cite{Aoyama:2020ynm,Aoyama:2012wk,Aoyama:2019ryr}, other choices for $\alpha$~\cite{Bouchendira_2011,Morel:2020dww} result in changes well within the quoted uncertainty.} 

\subsection{The EW contributions}

\begin{figure}[!t]
\centering
    \includegraphics[width= 0.8\textwidth]{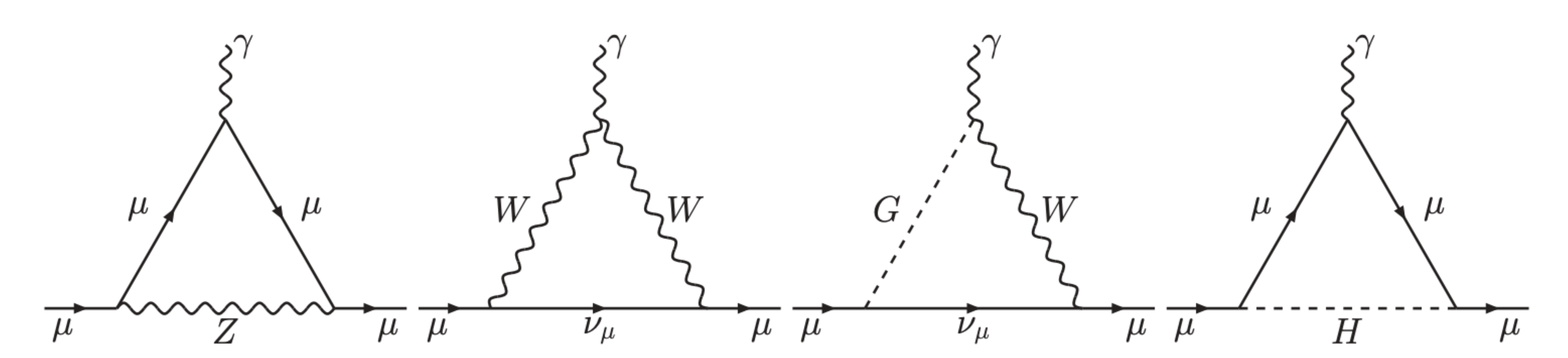}
  \caption{\small  One-loop EW Feynman diagrams. This figure is reprinted from~\cite{Aoyama:2020ynm}.}\label{fig:1loopEW} 
\end{figure}

The EW contributions constitute all diagrams that contain at least one of the EW bosons: $W$, $Z$, or Higgs (the one-loop Feynman diagrams are shown in Figure~\ref{fig:1loopEW}). As such, the EW contributions include QED and hadronic effects, but no EW processes enter the estimates of the QED, HVP, or HLbL sectors. Due to the masses of the EW bosons, the EW contributions to $a_\mu$ are highly suppressed. These contributions have been calculated up to two-loop and the three-loop contributions have been estimated~\cite{Czarnecki:2002nt,Gnendiger:2013pva}.

The value for $a_\mu^{\rm EW}$ is given as~\cite{Czarnecki:2002nt,Gnendiger:2013pva,Aoyama:2020ynm}
\begin{equation} \label{eq:alEW}
a_{\mu}^{\rm EW} =  153.6(1.0) \times 10^{-11} \, ,
\end{equation}
where the uncertainty includes the two-loop hadronic effects, neglected two-loop terms, and unknown three-loop contributions. The non-perturbative hadronic insertions that enter at two-loop significantly dominate the uncertainty of the EW contributions but, due to the EW suppression, are small relative to the HVP or HLbL sector uncertainties.

\subsection{The HVP contributions}

In general, the HVP contributions (depicted in the third diagram in Figure~\ref{fig:SMdiagrams}) can be calculated from data-driven approaches, utilizing measured $e^+e^-\rightarrow {\rm hadrons}$ data\footnote{Historically, the HVP contributions have also been computed using data from hadronic $\tau$ decays. However, the isospin-breaking corrections to the $\tau$ data are currently not at the understanding required to also be used as input into dispersion relations with the $e^+e^-\rightarrow {\rm hadrons}$ data (see~\cite{Aoyama:2020ynm} for further details).} (with or without applying additional constraints from e.g., analyticity) as input into dispersion relations, or from Lattice QCD. 

\subsection*{Data-driven HVP}

\begin{figure}[!t] 
\centering
\subfloat[LO.]{%
\includegraphics[width= 0.25\textwidth]{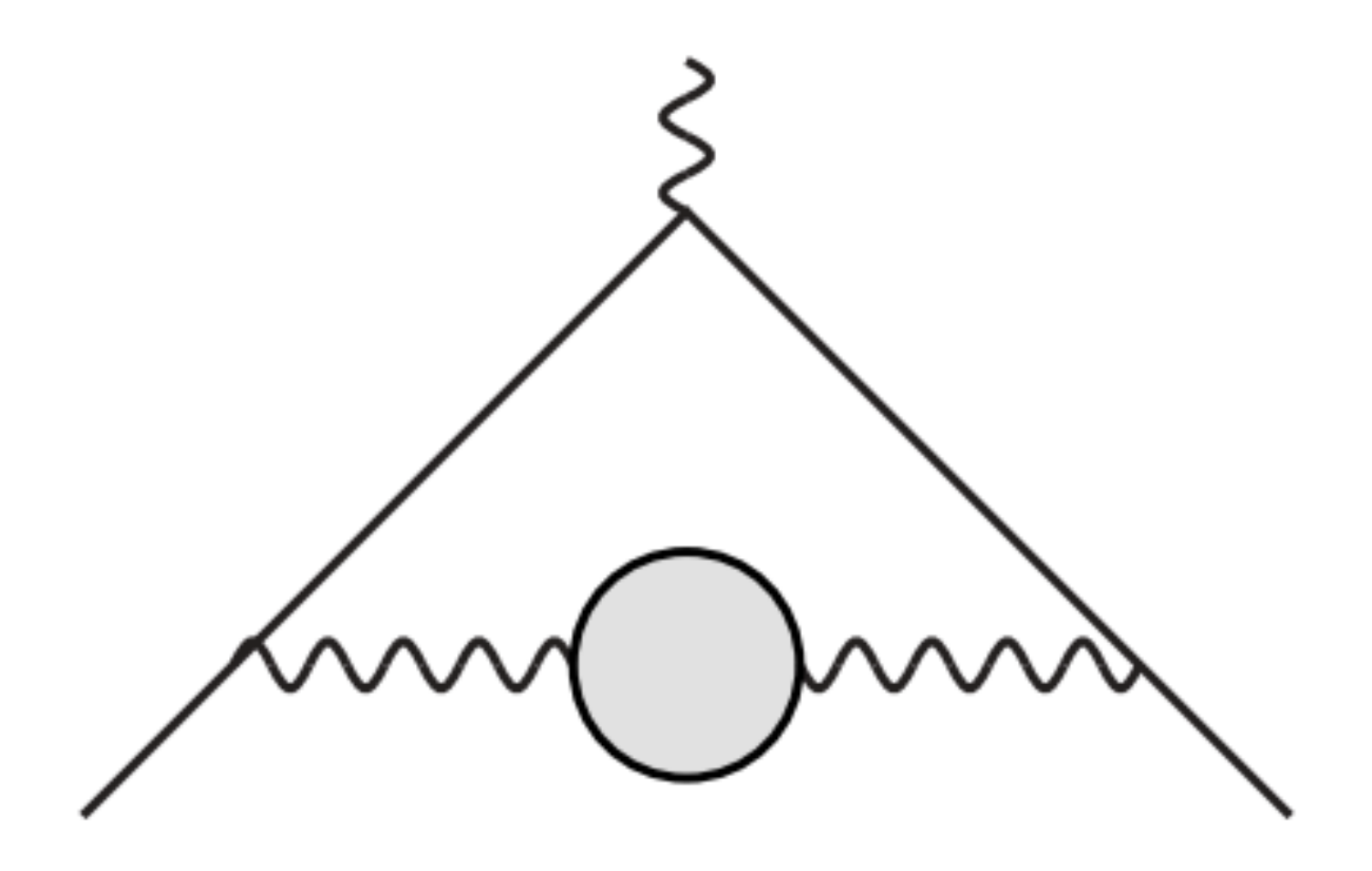}\label{fig:HVPLO}}\hfill
\subfloat[NLO.] {%
\includegraphics[width= 0.75\textwidth]{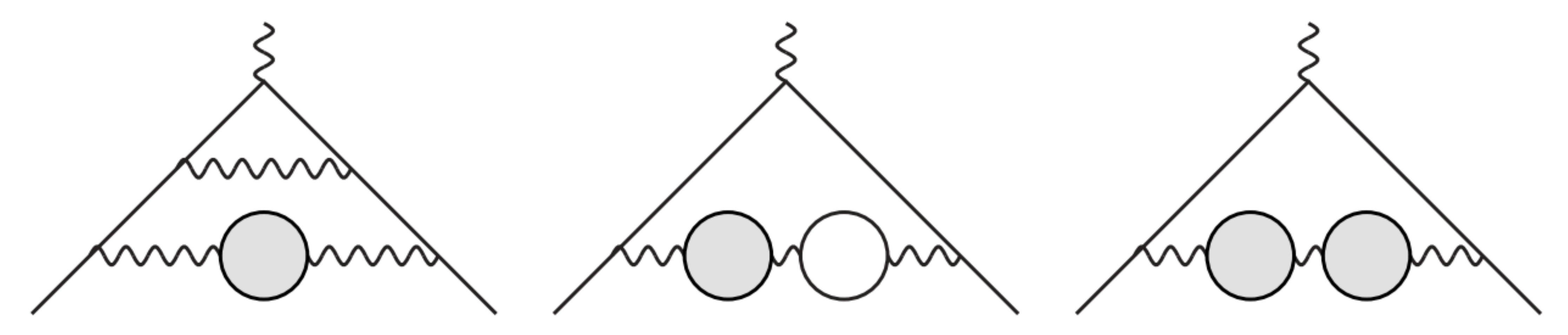}\label{fig:HVPNLO}}\hfill
  \caption{HVP Feynman diagrams at LO and NLO. The gray blobs refer to hadronic and the white ones in diagram (b) refer to leptonic VP. This figure is adapted from~\cite{Aoyama:2020ynm}.}
  \label{fig:HVP}
\end{figure}

The HVP contributions can be determined using a compilation of all available $e^+e^- \rightarrow {\rm hadrons}$ cross section data, $\sigma_{\rm had}(s) \equiv \sigma^0\left(e^+e^-\rightarrow \gamma^* \rightarrow {\rm hadrons} + \left(\gamma\right)\right) $, which is inclusive of final state radiation effects and where the superscript `0' indicates the cross section is bare (excluding all vacuum polarization effects). The leading order (LO) HVP contribution, shown in Figure~\ref{fig:HVPLO}, is evaluated from inputting the combined cross section data into the dispersion relation:
\begin{equation} 
a_{\mu}^{\rm LO\,HVP} =\frac{1}{4\pi^3}\int^{\infty}_{s_{th}} {\rm d}s \, K(s)\, \sigma_{\rm had}(s) \,,
\end{equation}
where $s_{th} = m_{\pi^0}^2$. $K(s)$ is a well-known kernel function~\cite{Brodsky:1967sr,Lautrup:1969fr} which gives greater weight to contributions from lower energies. Next-to-leading order (NLO) contributions are shown in Figure~\ref{fig:HVPNLO}. These and the NNLO contributions are determined from similar dispersion integrals and kernel functions with a data input identical to the LO case~\cite{Krause:1996rf}. 

\begin{figure}[!t]
\centering
    \includegraphics[width= 0.8\textwidth]{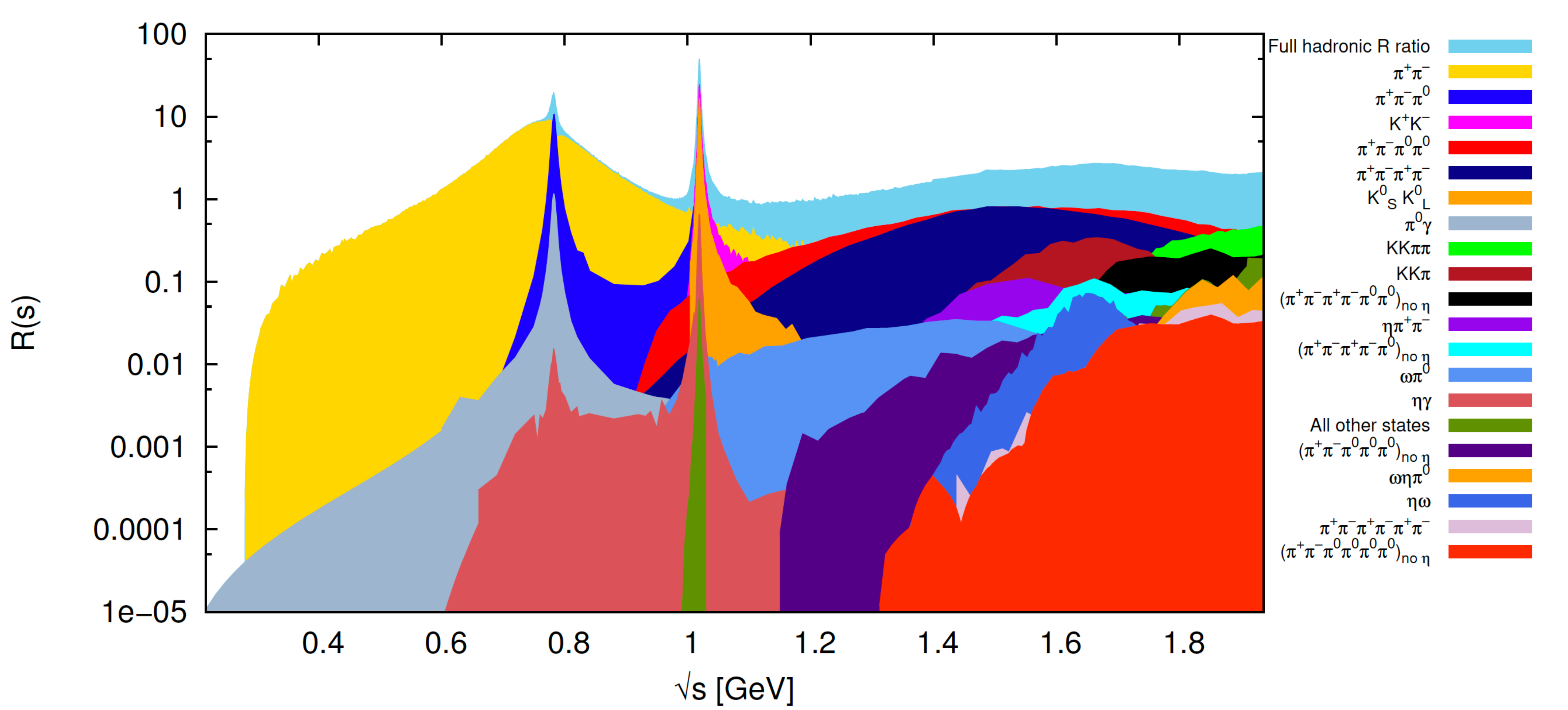}
  \caption{\small Contributions to the total hadronic cross section (expressed as the hadronic $R$-ratio, $R(s) = \sigma_{\rm had}(s)/\left(4\pi\alpha^2/\left(3s\right)\right)$) from the different final states below $\sqrt{s}\sim2$ GeV. The total hadronic cross section is shown in light blue and each final state is included as a new layer on top in decreasing order of the size of its contribution to $a_{\mu}^{\rm LO \, HVP}$. This figure has been taken from~\cite{Keshavarzi:2018mgv}.}\label{fig:KNTexclusive} 
\end{figure}

Numerous measurements for more than 35 exclusive hadronic channels (final states) from different
experiments must be combined, as depicted in Figure~\ref{fig:KNTexclusive}. Hadronic cross-section data are either obtained from direct scan measurements (e.g., CMD-2, SND, KEDR) or via the method of radiative return (e.g BaBar, KLOE, BES-III).\footnote{A detailed summary of the available hadronic cross section data and the experiments that measure them is given in~\cite{Aoyama:2020ynm}.} The data combinations are performed channel-by-channel to determine individual contributions to $a_\mu^{\rm HVP}$, which are then summed. The combinations themselves are non-trivial, with the combined result needing to be an accurate representation of the differing data and their uncertainties. The most dominant channel is the two pion channel, which contributes more than 70\% of the total HVP. 
Final states, threshold contributions, or resonances for which there are no data (but which are not negligible) are safely estimated. The estimated contribution of missing channels amounts to less than $0.05\%$ of $a_{\mu}^{\rm HVP}$~\cite{Davier:2017zfy,Davier:2019can,Aoyama:2020ynm}.

There are several data-driven different evaluations of $a_\mu^{\rm HVP}$ that differ in the treatment of the data as well as the assumptions made on the functional form of the cross section. The DHMZ~\cite{Davier:2017zfy,Davier:2019can} and KNT~\cite{Keshavarzi:2018mgv,Keshavarzi:2019abf} groups directly use the bare cross section. 
An alternative approach is achieved by the CHHKS groups, who apply additional constraints from analyticity and unitarity to evaluate the $\pi^0\gamma$, $2\pi$ and $3\pi$ channels ~\cite{Hoid:2020xjs,Colangelo:2018mtw,Hoferichter:2019gzf} (DHMZ apply similar constraints for the $2\pi$ channel~\cite{Davier:2019can}). The results from these three groups have been merged in a conservative procedure to account for differences between groups and tensions between data sets.
The merging procedure yields $a_{\mu}^{\rm LO\,HVP} = 6931(40) \times 10^{-11}$~\cite{Aoyama:2020ynm}, with the corresponding results for the $a_{\mu}^{\rm NLO\,HVP} = -98.3(7) \times 10^{-11}$~\cite{Keshavarzi:2019abf} and $a_{\mu}^{\rm NNLO\,HVP} = 12.4(1) \times 10^{-11}$~\cite{Kurz:2014wya} resulting in an estimate of the total HVP contribution of~\cite{Aoyama:2020ynm}
\begin{equation} \label{eq:amuHVP}
a_{\mu}^{\rm HVP} = 6845(40) \times 10^{-11} \, .
\end{equation}
A comprehensive review of all data-driven determinations of $a_{\mu}^{\rm HVP}$, including those from other groups not included in the merged result, is given in~\cite{Aoyama:2020ynm}.
Prospects to improve the data-driven determinations of $a_{\mu}^{\rm HVP}$ rest in new $e^+e^-\rightarrow{\rm hadrons}$ cross section measurements, particularly those of the $\pi^+\pi^-$ channel. Such new $\pi^+\pi^-$ data sets are expected from CMD-3~\cite{CMD-32piNew}, BaBar~\cite{BaBar2piNew}, BES-III~\cite{Ablikim:2019hff} and Belle-II~\cite{Belle-II2piNew}. The CMD-3 result is projected to be the most statistically precise of all the current measurements in the two pion channel, with systematic uncertainties ranging from 0.6\%-1\%. 

The same $\sigma_{\rm had}(s)$ data used to evaluate $a_{\mu}^{\rm HVP}$ are also used to estimate the five-flavor hadronic contribution to the running QED coupling at the $Z$-pole, $\Delta\alpha_{\rm had}^{(5)}(M_{Z}^2)$. This quantity is a crucial input to global EW fits and, therefore, predictions of the EW fit parameters (e.g., the Higgs mass, $m_H$). This connection has been explored in several works~\cite{Passera:2008jk,Crivellin:2020zul,Keshavarzi:2020bfy,deRafael:2020uif,Malaescu:2020zuc}, asking the following question: should the muon $g-2$ discrepancy be artificially accounted for in $\sigma_{\rm had}(s)$, are the predictions for EW fit parameters such as $m_H$ still consistent with their measured values? In~\cite{Keshavarzi:2020bfy}, shifts in $\sigma_{\rm had}(s)$ needed to bridge $\Delta a_{\mu}$ were found to be excluded above $\sqrt{s} \sim 0.7$~GeV at the 95\%CL. However, prospects for $\Delta a_{\mu}$ originating below that energy were deemed improbable given the required increases in the hadronic cross section. 
 
 Further opportunities to scrutinize the HVP contributions are expected from the MUonE experiment~\cite{Calame:2015fva,Abbiendi:2016xup,Abbiendi:2677471}, which is a proposed approach to determine the leading hadronic corrections to the muon $g-2$ purely from experiment. It proceeds by scattering high energy muons on atomic electrons of a low-Z target through the elastic process $\mu e \rightarrow \mu e$~\cite{Abbiendi:2016xup}. In doing so, it is possible to directly measure the running of the QED coupling, $\Delta\alpha(Q^2)$, for spacelike $Q^2$ (which is in common with lattice QCD determinations) and therefore extract the hadronic component $\Delta\alpha_{\rm had}(Q^2)$, which can be used as input into an alternative dispersion relation to give $a_{\mu}^{\rm LO\,HVP}$. Such an experimental result would serve as an invaluable cross check of the results from $e^+e^-\rightarrow {\rm hadrons}$ data and from lattice QCD. The current status for theory predictions, recent activities and future plans related to the proposed MUonE experiment is given in~\cite{Banerjee:2020tdt}.

\subsection*{HVP from Lattice QCD}

Determining the HVP contribution from lattice QCD is achieved by applying Euclidean spacetime discretization of the vacuum polarization tensor $\Pi_{\mu\nu}(Q^2)$ for spacelike $Q^2$ in finite volumes and with finite lattice spacing, which is then taken to continuum and infinite-volume limits.
In general, intermediate steps of these calculations can be chosen to be performed in different orders, resulting in differing intermediate results between analysis groups. Common to all determinations is the ability to split the calculation of $a_{\mu}^{\rm LO\,HVP}$ at $\mathcal{O}(\alpha^2)$ according to quark-connected (conn) and quark-disconnected (disc) contributions as 
\begin{equation}
     a_{\mu}^{\rm LO\,HVP}(\alpha^2) = a_{\mu,{\rm conn}}^{\rm LO\,HVP}  +  a_{\mu,{\rm disc}}^{\rm LO\,HVP}\,.
\end{equation}
As the different quark flavor-connections result in different statistical and systematic uncertainties, it is appropriate to separate them as
\begin{equation}
     a_{\mu,{\rm conn}}^{\rm LO\,HVP}  = a_{\mu}^{\rm LO\,HVP}(ud) + a_{\mu}^{\rm LO\,HVP}(s) + a_{\mu}^{\rm LO\,HVP}(c) + a_{\mu}^{\rm LO\,HVP}(b)\,,
     \label{eq:HLO}
\end{equation}
where $ud$ corresponds to the contributions of the light $u$ and $d$ quarks (treated in the isosymmetric limit $m_u = m_d$) and $s$, $c$ and $b$ are the strange, charm and bottom quark contributions, respectively. Modern lattice determinations (such as those described here) necessarily include strong and electromagnetic isospin-breaking corrections $\delta a_{\mu}^{\rm LO\,HVP}$ as $a_{\mu}^{\rm LO\,HVP} = a_{\mu}^{\rm LO\,HVP}(\alpha^2) + \delta a_{\mu}^{\rm LO\,HVP}$ .
The determinations of the isosymmetric flavor terms given in equation~\eqref{eq:HLO} and the corrections $\delta a_{\mu}^{\rm LO\,HVP}$ are prescription and scheme dependent, leading to differing and comparable results between lattice calculations. Although a full discussion of the various methods and analysis choices is beyond the scope of this review, a comprehensive examination is given in~\cite{Aoyama:2020ynm}. In general, however, all lattice prescriptions have common features. All results are extrapolated to the continuum and infinite-volume limits and interpolated or extrapolated to the physical point. Quoted errors contain both statistical and systematic uncertainties, where the systematics arise from common issues faced by all analyses: long-distance effects, finite-volume effects, discretization effects, scale setting, chiral extrapolation/interpolation and quark mass tuning. 

\begin{figure}[!t]
\centering
    \includegraphics[width= 0.6\textwidth]{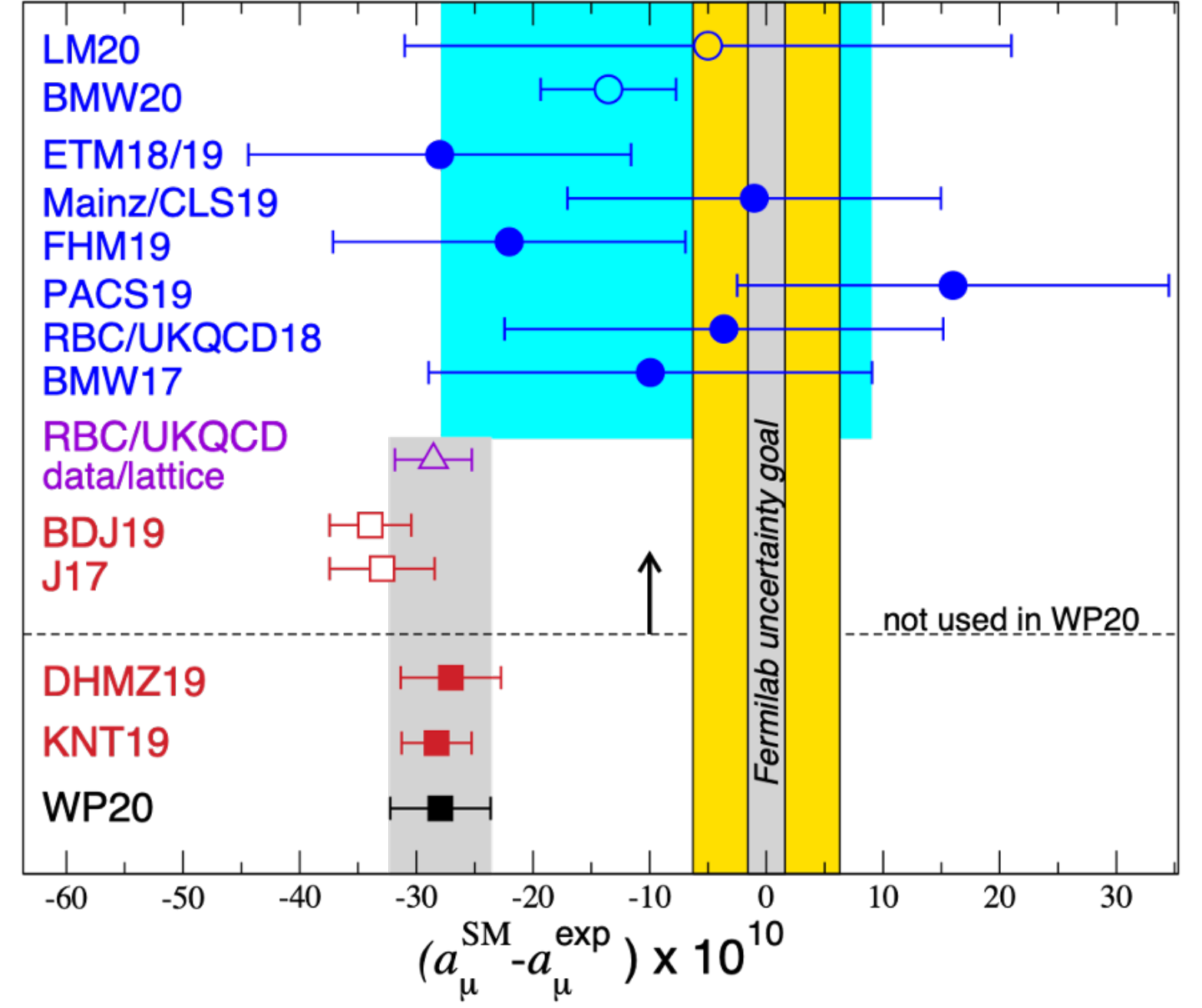}
  \caption{\small Results for $(a_{\mu}^{\rm SM} - a_{\mu}^{\rm EXP})\times10^{10}$ when $a_{\mu}^{\rm LO\,HVP}$ is taken from various lattice~\cite{Lehner:2020crt,Borsanyi:2020mff,Giusti:2018mdh,Giusti:2019xct,Gerardin:2019rua,Davies:2019efs,Chakraborty:2017tqp,Shintani:2019wai,Blum:2018mom,Borsanyi:2017zdw,DellaMorte:2017dyu,Chakraborty:2016mwy,Aubin:2019usy} and data-driven~\cite{Davier:2019can,Keshavarzi:2019abf,Jegerlehner:2017gek,Benayoun:2019zwh}. The filled dark blue circles are lattice results~\cite{Giusti:2018mdh,Giusti:2019xct,Gerardin:2019rua,Davies:2019efs,Chakraborty:2017tqp,Shintani:2019wai,Blum:2018mom,Borsanyi:2017zdw} that are included in the lattice average advocated in~\cite{Aoyama:2020ynm}, which is given indicated by the light-blue band. The unfilled dark blue circles are those results not included in the average~\cite{Lehner:2020crt,Borsanyi:2020mff,DellaMorte:2017dyu,Chakraborty:2016mwy,Aubin:2019usy,Burger:2013jya}. The red squares show results from data-driven determinations of $a_{\mu}^{\rm LO\,HVP}$, where filled squares are those included in the merged data-driven result~\cite{Davier:2019can,Keshavarzi:2019abf} (given by the black square marker and grey band) and unfilled squares are not~\cite{Jegerlehner:2017gek,Benayoun:2019zwh}. The purple triangle shows a hybrid result where noisy lattice data at very short and long distances are replaced by $e^+e^- \rightarrow {\rm hadrons}$ cross section data~\cite{Blum:2018mom}. The yellow band indicates the “no new physics” scenario, where $a_{\mu}^{\rm LO\,HVP}$ results are large enough to bring the SM prediction of $a_{\mu}$ into agreement with experiment. The grey band in the center of this indicates the projected experimental uncertainty from the Fermilab Muon $g-2$ experiment. This figure has been adapted from~\cite{Aoyama:2020ynm,AidaGm2TheorySeminar}.}\label{fig:latticeHVP} 
\end{figure}

Results from various lattice groups of the different flavor contributions and the total estimate of  $a_{\mu}^{\rm LO\,HVP}$ are given in Table~\ref{tab:latticeHVP} and shown in Figure~\ref{fig:latticeHVP}.\footnote{The results for $a_\mu^{\rm HVP}$ in this lattice HVP subsection are presented in units $10^{-10}$ owing to the generally lower precision of the quoted results. All other results for $a_\mu$ in this review are in units $10^{-11}$.} The ETM18/19~\cite{Giusti:2018mdh,Giusti:2019xct}, Mainz/CLS-19~\cite{Gerardin:2019rua}, FHM-19~\cite{Davies:2019efs,Chakraborty:2017tqp}, PACS-19~\cite{Shintani:2019wai}, RBC/UKQCD-18~\cite{Blum:2018mom} and BMW-17~\cite{Borsanyi:2017zdw} results are combined using a conservative procedure into an average of lattice results in~\cite{Aoyama:2020ynm}, which has value of
\begin{equation}
\label{eq:amuhlo_IB_lat}
    a_{\mu}^{\rm LO\,HVP} = 711.6(18.4) \times 10^{-10}\ ,
\end{equation}
and is shown by blue band in Figure~\ref{fig:latticeHVP}. These results have
values which span the range between the data-driven approaches and a no-new-physics scenario (shown by the green band), but generally with errors too large to make a definitive statement. Consequently, the error on the average is also consistent with both the data-driven approaches and the no-new-physics scenario. More recently, two results from the LM-20 and BMW-20 analyses found $a_{\mu}^{\rm LO\,HVP}[{\rm LM}$-${20}] = 714(30) \times 10^{-10}$~\cite{Lehner:2020crt} and $a_{\mu}^{\rm LO\,HVP}[{\rm BMW}$-${20}] = 707.5(5.5) \times 10^{-10}$~\cite{Borsanyi:2020mff} that are not included in the average of lattice results presented in~\cite{Aoyama:2020ynm}. The latter is the first lattice result for $a_{\mu}^{\rm LO\,HVP}$ with sub-percent precision. It is $1.3\sigma$ below the no-new-physics scenario and $2.1\sigma$ higher than the recommended data-driven result in equation~\eqref{eq:amuHVP}. Should the projected improvement in precision of other lattice evaluations confirm this result, then this difference must be understood.
\begin{table}[!t]
\begin{center}
\small
\begin{tabular}{lccccc}
\hline
\addlinespace[0.1cm]
Collaboration&\hspace{-0mm}
$a_{\mu}^{\rm LO\,HVP}(ud)$&\hspace{-0mm}
$a_{\mu}^{\rm LO\,HVP}(s)$&\hspace{-0mm}
$a_{\mu}^{\rm LO\,HVP}(c)$&\hspace{-0mm}
$a_{\mu,{\rm disc}}^{\rm LO\,HVP}$&\hspace{-0mm}
$a_{\mu}^{\rm LO\,HVP}$\\[0.1cm]
\hline
\addlinespace[0.05cm]
LM-20~\cite{Lehner:2020crt}&\hspace{-0mm}
657(29)&\hspace{-0mm}
52.8(7)&\hspace{-0mm}
14.3(7)&\hspace{-0mm}
$-11.2(4.0)$&\hspace{-0mm}
714\,(30)\\
BMW-20~\cite{Borsanyi:2020mff}&\hspace{-0mm}
633.7(4.7)&\hspace{-0mm}
53.4(1)&\hspace{-0mm}
14.6(1)&\hspace{-0mm}
$-18.6(2.0)$&\hspace{-0mm}
707.5\,(5.5)\\
\addlinespace[0.05cm]
\hdashline
\addlinespace[0.05cm]
ETM-18/19~\cite{Giusti:2018mdh,Giusti:2019xct}&\hspace{-0mm}
629.1(13.7)&\hspace{-0mm}
53.1(2.6)&\hspace{-0mm}
14.75(56)&\hspace{-0mm}
-&\hspace{-0mm}
692.1\,(16.3)\\
Mainz/CLS-19~\cite{Gerardin:2019rua}&\hspace{-0mm}
674(13) &\hspace{-0mm}
54.5(2.5)&\hspace{-0mm}
14.66(45)&\hspace{-0mm}
$-23.2(5.0)$&\hspace{-0mm}
720.0\,(15.9)\\
FHM-19~\cite{Davies:2019efs,Chakraborty:2017tqp}&\hspace{-0mm}
637.8(8.8) &\hspace{-0mm}
-&\hspace{-0mm}
-&\hspace{-0mm}
$-13(5)$&\hspace{-0mm}
699\,(15)\\
PACS-19~\cite{Shintani:2019wai}&\hspace{-0mm}
673(14) &\hspace{-0mm}
52.1(5)&\hspace{-0mm}
11.7(1.6)&\hspace{-0mm}
-&\hspace{-0mm}
$737\,(^{+15}_{-20})$\\
RBC/UKQCD-18~\cite{Blum:2018mom}&\hspace{-0mm}
649.7(15.0) &\hspace{-0mm}
53.2(5)&\hspace{-0mm}
14.3(7)&\hspace{-0mm}
$-11.2(4.0)$&\hspace{-0mm}
717.4\,(18.7)\\
BMW-17~\cite{Borsanyi:2017zdw}&\hspace{-0mm}
647.6(19.2) &\hspace{-0mm}
53.73(49)&\hspace{-0mm}
14.74(16)&\hspace{-0mm}
$-12.8(1.9)$&\hspace{-0mm}
711.1\,(19.0)\\
\addlinespace[0.05cm]
\hdashline
\addlinespace[0.05cm]
Mainz/CLS-17~\cite{DellaMorte:2017dyu}&\hspace{-0mm}
588.2(35.8) &\hspace{-0mm}
51.1(1.7)&\hspace{-0mm}
14.3(2)&\hspace{-0mm}
-&\hspace{-0mm}
$654\,(^{+38}_{-39})$\\
HPQCD-16~\cite{Chakraborty:2016mwy}&\hspace{-0mm}
$599.0(12.5)$&\hspace{-0mm}
-&\hspace{-0mm}
-&\hspace{-0mm}
$0(9)(-)$&\hspace{-0mm}
667\,(14)\\
\addlinespace[0.05cm]
\hline
\end{tabular}
\caption{\small Lattice results for flavor-specific ($ud$,$s$,$c$,disc) contributions and the full evaluations of $a_{\mu}^{\rm LO\,HVP}$~\cite{Lehner:2020crt,Borsanyi:2020mff,Giusti:2018mdh,Giusti:2019xct,Gerardin:2019rua,Davies:2019efs,Chakraborty:2017tqp,Shintani:2019wai,Blum:2018mom,Borsanyi:2017zdw,DellaMorte:2017dyu,Chakraborty:2016mwy,Aubin:2019usy}. Where available, these results include all necessary corrections (e.g., strong and QED isospin breaking corrections), although these corrections are not explicitly stated here. The errors displayed are the statistical and systematic uncertainties, added in quadrature. All results are given as $a_{\mu}^{\rm LO\,HVP}\times10^{10}$. The ETM18/19~\cite{Giusti:2018mdh,Giusti:2019xct}, Mainz/CLS-19~\cite{Gerardin:2019rua}, FHM-19~\cite{Davies:2019efs,Chakraborty:2017tqp}, PACS-19~\cite{Shintani:2019wai}, RBC/UKQCD-18~\cite{Blum:2018mom} and BMW-17~\cite{Borsanyi:2017zdw} results in-between the dashed lines are those included in the average of lattice results advocated in~\cite{Aoyama:2020ynm}.}
\label{tab:latticeHVP}
\end{center}
\end{table}

In general, continued progress in the precision of the wider set of lattice evaluations of $a_{\mu}^{\rm LO\,HVP}$ is expected~\cite{Aoyama:2020ynm}. The major challenges in reducing the uncertainties come from finite-volume effects, exponentially growing signal-to-noise problems at large Euclidean times, disconnected contributions, and strong isospin breaking and QED corrections. Nonetheless, the prospects of reducing both the statistical and systematic uncertainties of lattice evaluations of the hadronic vacuum polarization appear unhindered and extremely likely to be achieved in the near future.

\subsection{The HLbL contributions}

\begin{figure}[!t]
\centering
    \includegraphics[width= 0.4\textwidth]{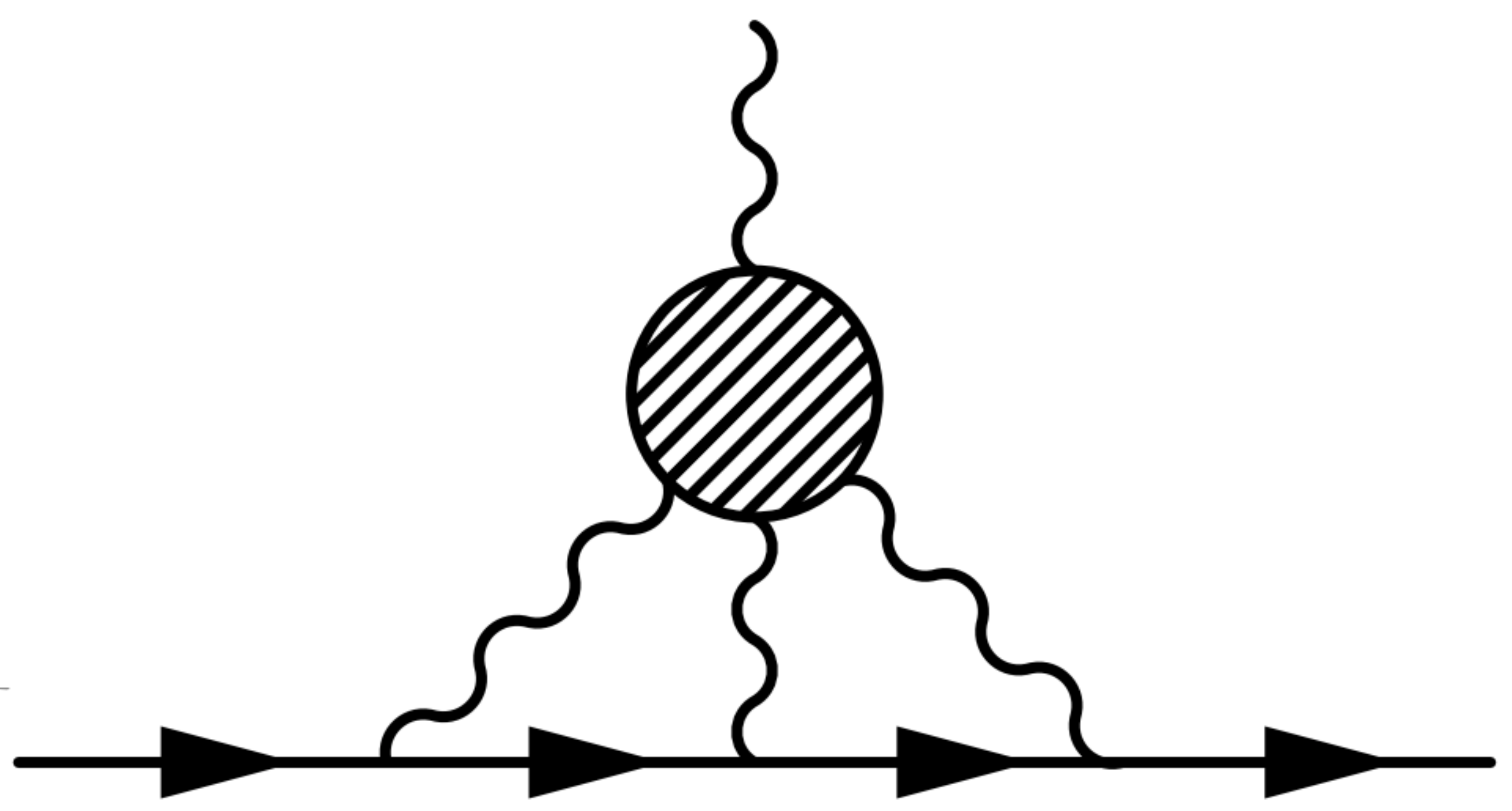}
  \caption{\small  HLbL Feynman diagram at LO. The shaded blob represents all possible intermediate hadronic states. This figure is reprinted from~\cite{Aubin:2019usy}.}\label{fig:HLbL} 
\end{figure}

Contributions from HLbL scattering, shown in Figure~\ref{fig:HLbL} and in the furthest-right diagram of Figure~\ref{fig:SMdiagrams}, describe the process whereby an external soft and on-shell photon interacts through a hadronic blob with three off-shell photons that then couple
to the muon. As such, they are classified by a four-point function and therefore require calculations that are more complicated than those of the two-point HVP function. The HLbL contributions enter at $\mathcal{O}(\alpha^3)$ and are consequently suppressed by an additional order of $\alpha$, making them two orders of magnitude smaller than the vacuum polarization sector. The hadronic contributions to LbL scattering arise from single mesons (e.g $\pi^0,\eta,\eta',f_0(980),\allowbreak a_0(980)$), axial-vector mesons (e.g., $a_1,f_1$), tensor mesons (e.g., $f_2,a_2$) and charged pion or kaon loops. They have, in the past, been determined through model-dependent estimates from meson exchanges, the large $N_c$ limit, chiral perturbation theory estimates, short distance constraints from the operator product expansion and pQCD. Over time, several different approaches to evaluating $a_{\mu}^{\rm had, \, LbL}$ have been attempted~\cite{Melnikov:2003xd,Prades:2009tw,Nyffeler:2009tw, Jegerlehner:2009ry,Jegerlehner:2017gek}, resulting in good agreement for the leading $N_c$ ($\pi^0$ exchange) contribution, but differing for sub-leading effects. The widely accepted model-based estimates of $a_{\mu}^{\rm HLbL}$ were $a_{\mu}^{\rm HLbL} [{\rm PdRV}(09)] = 105(26)
\times 10^{-11}$~\cite{Prades:2009tw}, $a_{\mu}^{\rm HLbL} [{\rm N/JN(09)}] = 116(39)\times 10^{-11}$~\cite{Jegerlehner:2009ry} and $a_{\mu}^{\rm HLbL} [{\rm J(17)}] = 100.4(28.2)\times 10^{-11}$~\cite{Jegerlehner:2017gek}. The uncertainties of these evaluations, as well as being larger than the required precision to improve the prediction of $a_\mu^{\rm SM}$, were model-dependent and therefore often difficult to confidently quantify. Fortunately, motivated by the muon $g-2$ theory initiative~\cite{Aoyama:2020ynm}, the status of the determinations of $a_{\mu}^{\rm had, \, LbL}$ are now vastly improved due to newer calculations from data-driven dispersive approaches and from lattice QCD.

\subsubsection*{Data-driven and dispersive HLbL}

Modern data-driven and dispersive methods to calculate $a_{\mu}^{\rm HLbL}$ provide a model-independent evaluation, as well as there being several major improvements in the model-dependent estimates for sub-leading contributions~\cite{Masjuan:2017tvw,Colangelo:2017fiz,Hoferichter:2018kwz,Gerardin:2019vio,Bijnens:2019ghy,Colangelo:2019uex,Pauk:2014rta,Danilkin:2016hnh,Knecht:2018sci,Eichmann:2019bqf,Roig:2019reh}. Where possible, experimental data are used as input for various hadronic insertions and, if data are not available, theoretical calculations of the amplitudes are employed. 
The HLbL tensor can be unambiguously split into the sum of all intermediate states in direct and crossed channels as $\Pi_{\mu\nu\lambda\sigma} = \Pi_{\mu\nu\lambda\sigma}^{\pi^0\text{-pole}} + \Pi_{\mu\nu\lambda\sigma}^{\pi\text{-box}} +  \Pi_{\mu\nu\lambda\sigma}^{\pi\pi} + \ldots$ It follows that 
\begin{equation}
a_{\mu}^{\rm HLbL} = a_\mu^{\pi^0\text{-pole}} + a_\mu^{\pi\text{-box}} +  a_\mu^{\pi\pi} + \ldots\, 
\end{equation}
$a_{\mu}^{\rm HLbL}$ is understood to be dominated by contributions originating below $1.5\,{\rm{GeV}}$, with the $\pi^0$-pole being the most dominant contribution. Other single-particle states (e.g., $\eta$ and $\eta'$) are suppressed, as are two pion and two kaon effects further still, although it should be noted that adapting these expressions for $\eta$, $\eta'$ and two-kaon states is straightforward. The major experimental inputs to evaluate these expressions come in the form of $\pi$, $\eta$ and $\eta'$ transition form factors (TFFs). These data come in various forms as either single-virtual TFFs in the spacelike regime from $\gamma^*\gamma$ collisions or in the timelike region from radiative production in $e^+e^-$ collisions, single or double Dalitz decays of pseudoscalars, double-virtual TFFs, or from Dalitz decays of vector mesons.

Detailed descriptions of the dispersive calculations and experimental inputs of each of the various contributions to $a_{\mu}^{\rm HLbL}$ are given in~\cite{Aoyama:2020ynm,Masjuan:2017tvw,Colangelo:2017fiz,Hoferichter:2018kwz,Gerardin:2019vio,Bijnens:2019ghy,Colangelo:2019uex,Pauk:2014rta,Danilkin:2016hnh,Knecht:2018sci,Eichmann:2019bqf,Roig:2019reh}. The recommended results of these calculations, compared with previous estimates, are displayed in Table~\ref{tab:compilations}. In general, the values obtained from dispersive approaches are consistent with those from previous estimates, with improved uncertainties. The sum of the values from the different contributions results in a data-driven, dispersive estimate for the full $a_{\mu}^{\rm HLbL}$ of
\begin{equation}\label{eq:dataHLbL}
a_{\mu}^{\rm HLbL} = 92(19) \times 10^{-11}~,
\end{equation}
where the overall uncertainty is established as a sum of data-driven errors added in quadrature and model-dependent errors added linearly~\cite{Aoyama:2020ynm}. The NLO HLbL contribution is found to be $a_{\mu}^{\rm NLO\, HLbL} = 2(1) \times 10^{-11}$. Further improvements on these dispersive estimates are currently in progress and are expected to achieve a 10\% uncertainty on the HLbL contribution to $a_\mu$~\cite{Aoyama:2020ynm}.
\begin{table}
\centering
\small 
\begin{tabular}{crrrr}
\toprule
  Contribution & \quad PdRV(09)~\cite{Prades:2009tw} \quad & 
  \quad N/JN(09)~\cite{Nyffeler:2009tw, Jegerlehner:2009ry} \quad & 
  \quad J(17)~\cite{Jegerlehner:2017gek} \quad & 
  \quad Dispersive~\cite{Aoyama:2020ynm} \quad
  \tabularnewline  
\midrule
 $\pi^0,\eta,\eta'$-poles & $ 114(13) $ & $ 99(16) $ & $ 95.45
 (12.40) $ & $ 93.8(4.0) $ 
\tabularnewline 

$\pi,K$-loops/boxes & $ -19(19) $ & $ -19(13) $ & $ -20(5) $ & 
 $ -16.4(2) $
\tabularnewline 
$S$-wave $\pi\pi$ rescattering & $ -7(7) $ & $ -7(2) $ & $-5.98(1.20) $ &  
 $ -8(1) $
\tabularnewline 
\midrule
subtotal & $88(24)$ & $73(21)$ & $69.5(13.4)$ & $69.4(4.1)$
\tabularnewline 
\midrule
scalars & $-$ & $-$ & $-$ &    
  \multirow{2}{*}{$\bigg\}\qquad -1(3)$} 
 \tabularnewline 

tensors & $-$ & $-$ & $ 1.1(1) $ & 
 \tabularnewline 

axial vectors & $ 15(10) $ & $ 22(5) $ & $ 7.55(2.71) $ &  
 $ 6(6) $ 
\tabularnewline 

~$u,d,s$-loops / short-distance~ & $-$ & $ 21(3) $ & $ 20(4) $
&   $ 15(10) $   
\tabularnewline \midrule
$c$-loop & $2.3$ & $-$ & $2.3(2)$
&   $ 3(1)$   
\tabularnewline 
\midrule
total & $ 105(26) $ & $ 116(39) $ & $ 100.4(28.2) $ &   
  $ 92(19) $ 
\tabularnewline 
\bottomrule
\end{tabular}
\caption{Results from various contributions to $a_{\mu}^{\rm HLbL}$ in units of $10^{-11}$ from dispersive evaluations~\cite{Aoyama:2020ynm} and compared with previous estimates~\cite{Prades:2009tw,Nyffeler:2009tw, Jegerlehner:2009ry,Jegerlehner:2017gek}. This table has been adapted from~\cite{Aoyama:2020ynm}.} 
  \label{tab:compilations}
\end{table}

\subsection*{HLbL from Lattice QCD}

Due to the efforts of the Muon $g-2$ Theory Initiative~\cite{Aoyama:2020ynm}, the full $a_{\mu}^{\rm HLbL}$ has now been calculated on the lattice by two groups~\cite{Blum:2019ugy,Chao:2021tvp}. In discretized Euclidean spacetime, it has been computed treating QED both perturbatively and non-perturbatively, in both finite (QED$_L$) and infinite volumes (QED$_\infty$). Large uncertainties arise from volume errors and non-zero lattice spacings. In QED$_L$, $a_{\mu}^{\rm HLbL}$ is recovered by extrapolating to infinite-volume and continuum limits. Derivations and the methodologies of the approaches are given in detail in~\cite{Aoyama:2020ynm,Blum:2019ugy,Chao:2021tvp}. In general, both approaches have been tested by replacing quark loops with lepton loops and have been shown to perform well. Additionally, cross checks have been performed between the results of both groups, which exhibit compatibly when checking effects from lattice spacings and finite/infinite volumes.

After the infinite volume and continuum extrapolations, the result from the RBC calculation (with both QED and QCD gauge fields on the finite-volume QED$_L$) found~\cite{Blum:2019ugy}
\begin{equation}\label{eqn:hlbllatrbcres}
a_{\mu}^{\rm HLbL} = 78.7(30.6)_\text{stat}(17.7)_\text{sys}\times 10^{-11} \, .
\end{equation}
This calculation was performed for several lattice ensembles, with different lattice spacing and volume, with all particles at their physical masses and including contributions from both connected and disconnected diagrams. The result is not currently as precise as the dispersive HLbL determination, but is expected to improve in precision for both the statistical and systematic errors. Until recently, this result was the only complete calculation of $a_{\mu}^{\rm HLbL}$ and so is the recommended value for $a_{\mu}^{\rm HLbL}$ given in~\cite{Aoyama:2020ynm}. In QED$_\infty$, the RBC group has carried out preliminary calculations of both connected and leading disconnected diagrams with physical masses. A more recent calculation from the Mainz group found $a_{\mu}^{\rm HLbL} = 107(15) \times 10^{-11}$~\cite{Chao:2021tvp}, which is consistent with the result from~\cite{Blum:2019ugy}, but with a smaller uncertainty. Further improved results of $a_{\mu}^{\rm HLbL}$ from both groups in the near future.

\subsection{The SM prediction for $a_\mu$}

The recommended value for the SM prediction of the muon's anomalous magnetic moment is~\cite{Aoyama:2020ynm}
\begin{eqnarray}
a^{\rm{SM}}_{\mu} =116~591~810(43) \times 10^{-11}~(0.37\,\rm{ppm})\, .
\label{eq:amuvalue}
\end{eqnarray}
The various contributions used to reach this value\footnote{We remind the reader that whenever the value in equation~\eqref{eq:amuvalue} is used, the following original references should be cited:~\cite{Aoyama:2012wk,Aoyama:2019ryr,Czarnecki:2002nt,Gnendiger:2013pva,Davier:2017zfy,Keshavarzi:2018mgv,Colangelo:2018mtw,Hoferichter:2019gzf,Davier:2019can,Keshavarzi:2019abf,Kurz:2014wya,Melnikov:2003xd,Masjuan:2017tvw,Colangelo:2017fiz,Hoferichter:2018kwz,Gerardin:2019vio,Bijnens:2019ghy,Colangelo:2019uex,Blum:2019ugy,Colangelo:2014qya}.} are summarized in Table~\ref{tab:summary}. The QED value is that given in equation~\eqref{eq:alQED} and the EW contribution is that given in equation~\eqref{eq:alEW}. For the HVP, as the lattice evaluations are at present not precise enough to be comparable to those from dispersive approaches, the recommended value is the merged result from data-driven analyses given in equation~\eqref{eq:amuHVP}. For the HLbL contributions, as the dispersive (equation~\eqref{eq:dataHLbL}) and lattice (equation~\eqref{eqn:hlbllatrbcres}) evaluations are in good agreement, a weighted average of the two values is used to find a LO + NLO result of $a_{\mu}^{\rm LO + NLO,\, HLbL} = 92(18)\times 10^{-11}$.

\begin{table}[!t]
\begin{centering}
\small
\begin{tabular}{l r l }
\hline
Contribution &  
Value $\times 10^{11}$ & 
References \\ 
\hline
HVP LO ($e^+e^-$)  &  $ 6931(40) $  &
~\cite{Davier:2017zfy,Keshavarzi:2018mgv,Colangelo:2018mtw,Hoferichter:2019gzf,Davier:2019can,Keshavarzi:2019abf} \\
HVP NLO ($e^+e^-$) & 
$-98.3(7)$ &
~\cite{Keshavarzi:2019abf}\\
HVP NNLO ($e^+e^-$) &
$12.4(1)$ &
~\cite{Kurz:2014wya}\\
HVP LO (lattice, $udsc$) & 
$7116 (184)$ & 
~\cite{Chakraborty:2017tqp,Borsanyi:2017zdw,Blum:2018mom,Giusti:2019xct,Shintani:2019wai,Davies:2019efs,Gerardin:2019rua,Aubin:2019usy,Giusti:2019hkz}\\
HLbL (phenomenology) & 
$92(19)$ & 
~\cite{Melnikov:2003xd,Masjuan:2017tvw,Colangelo:2017fiz,Hoferichter:2018kwz,Gerardin:2019vio,Bijnens:2019ghy,Colangelo:2019uex,Pauk:2014rta,Danilkin:2016hnh,Jegerlehner:2017gek,Knecht:2018sci,Eichmann:2019bqf,Roig:2019reh}\\
HLbL NLO (phenomenology) & 
$2(1)$ & 
~\cite{Colangelo:2014qya}\\
HLbL (lattice, $uds$) & 
$79(35)$ & 
~\cite{Blum:2019ugy}\\
HLbL (phenomenology + lattice) &
$90(17)$ & 
~\cite{Melnikov:2003xd,Masjuan:2017tvw,Colangelo:2017fiz,Hoferichter:2018kwz,Gerardin:2019vio,Bijnens:2019ghy,Colangelo:2019uex,Pauk:2014rta,Danilkin:2016hnh,Jegerlehner:2017gek,Knecht:2018sci,Eichmann:2019bqf,Roig:2019reh,Blum:2019ugy}\\
\hline
QED             &  
$116\,584\,718.931(104)$  & ~\cite{Aoyama:2012wk,Aoyama:2019ryr}\\
Electroweak     &  
$153.6(1.0)$   & 
~\cite{Czarnecki:2002nt,Gnendiger:2013pva}\\
HVP ($e^+e^-$, LO + NLO + NNLO) & 
$6845(40)$ & 
~\cite{Davier:2017zfy,Keshavarzi:2018mgv,Colangelo:2018mtw,Hoferichter:2019gzf,Davier:2019can,Keshavarzi:2019abf,Kurz:2014wya} \\
HLbL (phenomenology + lattice + NLO) & $92(18)$ & 
~\cite{Melnikov:2003xd,Masjuan:2017tvw,Colangelo:2017fiz,Hoferichter:2018kwz,Gerardin:2019vio,Bijnens:2019ghy,Colangelo:2019uex,Pauk:2014rta,Danilkin:2016hnh,Jegerlehner:2017gek,Knecht:2018sci,Eichmann:2019bqf,Roig:2019reh,Blum:2019ugy,Colangelo:2014qya}\\ 
\hline 
\hline
Total SM Value  & 
$ 116\, 591\, 810(43) $  & ~\cite{Aoyama:2012wk,Aoyama:2019ryr,Czarnecki:2002nt,Gnendiger:2013pva,Davier:2017zfy,Keshavarzi:2018mgv,Colangelo:2018mtw,Hoferichter:2019gzf,Davier:2019can,Keshavarzi:2019abf,Kurz:2014wya,Melnikov:2003xd,Masjuan:2017tvw,Colangelo:2017fiz,Hoferichter:2018kwz,Gerardin:2019vio,Bijnens:2019ghy,Colangelo:2019uex,Blum:2019ugy,Colangelo:2014qya}\\
        \bottomrule
	\end{tabular}
	\caption{Summary of the contributions to $a_\mu^{\rm SM}$. This table has been adapted from~\cite{Aoyama:2020ynm}.}
\label{tab:summary}
\end{centering}
\end{table}

\section{Constraints on BSM physics}

The new result from the Fermilab Muon $g-2$ experiment yields a value for $\Delta a_\mu$ that is larger than the EW contributions to $a_\mu$ and, therefore, provides a large constraint on several possible scenarios of physics beyond the SM (BSM). In this section, we provide an overview of these models and refer the interested readers to the references in~\cite{Crivellin:2019mvj, Athron:2021iuf} for a more robust and extremely detailed discussion of the BSM constraints coupled to $\Delta a_\mu$ reviewed here.

These models require some enhancement to be compatible with the large $\Delta a_\mu$ and include single field extensions of the SM (e.g., dark photon, dark $Z$, two-Higgs doublet model (2HDM), scalar leptoquarks), two-field extensions of the SM (e.g., vector-like leptons, scalar singlet plus fermion), three-field extensions of the SM (e.g. two/one scalars plus one/two charged fermions and scalar/fermionic dark matter), various supersymmetry (SUSY) scenarios and several other possibilities. Typically, models that resolve $\Delta a_\mu$ require new states with masses $\lesssim$ 1~TeV. However, connected results from the LHC and dark matter searches which have not found SM signals have led to tensions in many models explaining the $g-2$ discrepancy and have constrained the possibilities of certain BSM scenarios further still. The interplay with $a_e$ is also interesting. Should a negative $\Delta a_e$ be realized, the resulting model would need to induce lepton flavor decoupling to ensure overall contributions of different signs to $a_e$ and $a_\mu$. Dark photons, for example, lead to contributions of a positive sign in both and could therefore increase the tension in $a_e$. 

$a_\mu$ has properties that directly associate it with several SM observables and appropriate BSM models: it is CP-conserving, flavor conserving, loop induced, and chirality flipping~\cite{Athron:2021iuf}. The latter, in particular, opens it up to scenarios involving broken chiral symmetry, spontaneous breaking of EW gauge invariance, chirality enhancement, or invoking light new particles. BSM contributions to $a_\mu$ are suppressed by $1/M_{\rm BSM}^2$, where $M_{\rm BSM}$ is the BSM mass scale. Typically, these contributions also induce BSM loop-contributions to the muon mass $m_\mu$. For such scenarios, the upper limit for the BSM mass scale which can resolve $\Delta a_\mu$ is $M_{\rm BSM} \lesssim \mathcal{O}(2.1)$ TeV~\cite{Athron:2021iuf}. 

The analysis in~\cite{Athron:2021iuf} has performed a robust investigation of the possibility of SUSY explanations of $\Delta a_\mu$ when coupled with external constraints. In general, it was shown that continued lack of evidence for BSM physics from the LHC and dark matter searches have greatly restricted the SUSY parameter space and imply mass patterns. MSSM scenarios with heavy charginos and smuons are disfavored and can only explain $\Delta a_\mu$ if $\tan \beta >> 40$ and/or $\mu >> 4$~TeV. A Bino-like LSP is a promising candidate, with several viable areas of parameter space available that allow for Wino masses within LHC limits and ranging values of $\tan\beta$. Scenarios involving Wino- or Higgsino-like LSP can also accommodate $\Delta a_\mu$, with the  Wino-like scenario
leading to the largest possible MSSM contribution to $\Delta a_\mu$. Both these scenarios, however, require additional non-MSSM dark matter components (e.g., gravitinos). In other scenarios where both charginos are assumed to be lighter the sleptons are also limited by dark matter searches and LHC data. 

Very small amounts of parameter space remain for the 2HDM (which also preserves minimal flavor-violation) as a viable candidate for $\Delta a_\mu$, but only for lepton-specific type X or generalized flavor aligned versions. All other types of 2HDM with minimal flavor violation are excluded. Many leptoquark models are also excluded, with the LHC constraining leptoquark masses to be much greater than $1$~TeV such that $\Delta a_\mu$ explanations can lead to very large loop corrections to the muon mass which may be interpreted as fine-tuning~\cite{Athron:2021iuf}. 
Dark photon solutions for $\Delta a_\mu$ are largely excluded from results for the electron anomalous magnetic moment $a_e$ and various dark photon and dark matter searches. Dark $Z$ models (see Section~\ref{sec:darkZ}) such as those with $M_{Z_\mu} < 1$~GeV and predominant couplings to $\mu,\tau$-leptons through the gauged $L_{\mu}-L_{\tau}$ quantum numbers are less restricted, but still strongly constrained. Simple models introducing three new fields are the least restricted. Although they require large coupling constants, they can accommodate $a_\mu$ and dark matter over a large amount of parameter space. 

Note that chiral enhancement can additionally lead to a large muon (or electron) EDM, $d_\mu$ ($d_e$)~\cite{Crivellin:2019mvj}. The Muon $g-2$ experiments at Fermilab~\cite{Chislett:2016jau} and J-PARC~\cite{Abe:2019thb}, as well as a dedicated experiment at PSI~\cite{Adelmann:2021udj}, aim to improve the current limits on the muon EDM by 2-4 orders of magnitude (see Section~\ref{sec:muEDM}). Ref~\cite{Crivellin:2018qmi} describes three scenarios: (1) $\Delta a_e > 0$ and small,  $\Delta a_\mu > 0$: this scenario is allowed via minimal flavor violation but constrains $d_\mu$ beyond what is experimentally viable. (2) $\Delta a_e < 0$ and large,  $\Delta a_\mu > 0$: this scenario is possible through flavor-violating chiral enhancement and could result in a large muon EDM. (3) $\Delta a_e > 0$ and large,  $\Delta a_\mu > 0$: this scenario is allowed in models introducing light new particles such as the dark photon but would result in $d_\mu = 0$. These scenarios are additionally constrained by searches for charged lepton flavor violation (cLFV), for example the MEG bounds of the process $\mu\rightarrow e\gamma$~\cite{TheMEG:2016wtm}. The status of cLFV experiments is discussed in Section~\ref{sec:cLFV}.

\section{Future muon $g-2$ experiments}

\subsection{J-PARC E34 experiment}

The muon $g-2$/EDM collaboration at J-PARC (E34)~\cite{Abe:2019thb} in Japan aims to provide an independent measurement of $a_{\mu}$ with a completely new approach in terms of the muon beamline, storage ring conditions, and positron detection. The experiment will utilize a low-emittance 300~MeV/$c$ muon beam, which is produced by re-acceleration of thermal muons regenerated by the laser resonant ionization of muonium ($\mu^{+}e^{-}$, or Mu) atoms emitted from a silica aerogel. The use of a low emittance beam eliminates the need for strong focusing by an electric field, which introduces a large correction to the $\omega_{a}$ measurement and a sizable systematic error in the BNL and Fermilab experiments (the electric-field correction was $C_{e}=300-500$~ppb depending on run conditions while the uncertainty on the correction was $\sim 50$~ppb) and the muon momentum can therefore be selected different to the "magic momentum". The muon beam will be stored in a highly uniform 3.0~T magnetic storage ring that is 20 times smaller than the Fermilab storage ring. The energy or momentum of the decay positrons will be fully reconstructed using 40 radially arranged silicon strip sensors, rather than the calorimetry.
Phase-I of the experiment is expected to begin in 2025 with a target precision of about $0.5$~ppm, similar to that of BNL or the Run-1 result from Fermilab. 

\begin{figure}[htbp]
\centering
    \includegraphics[width= 0.8\textwidth]{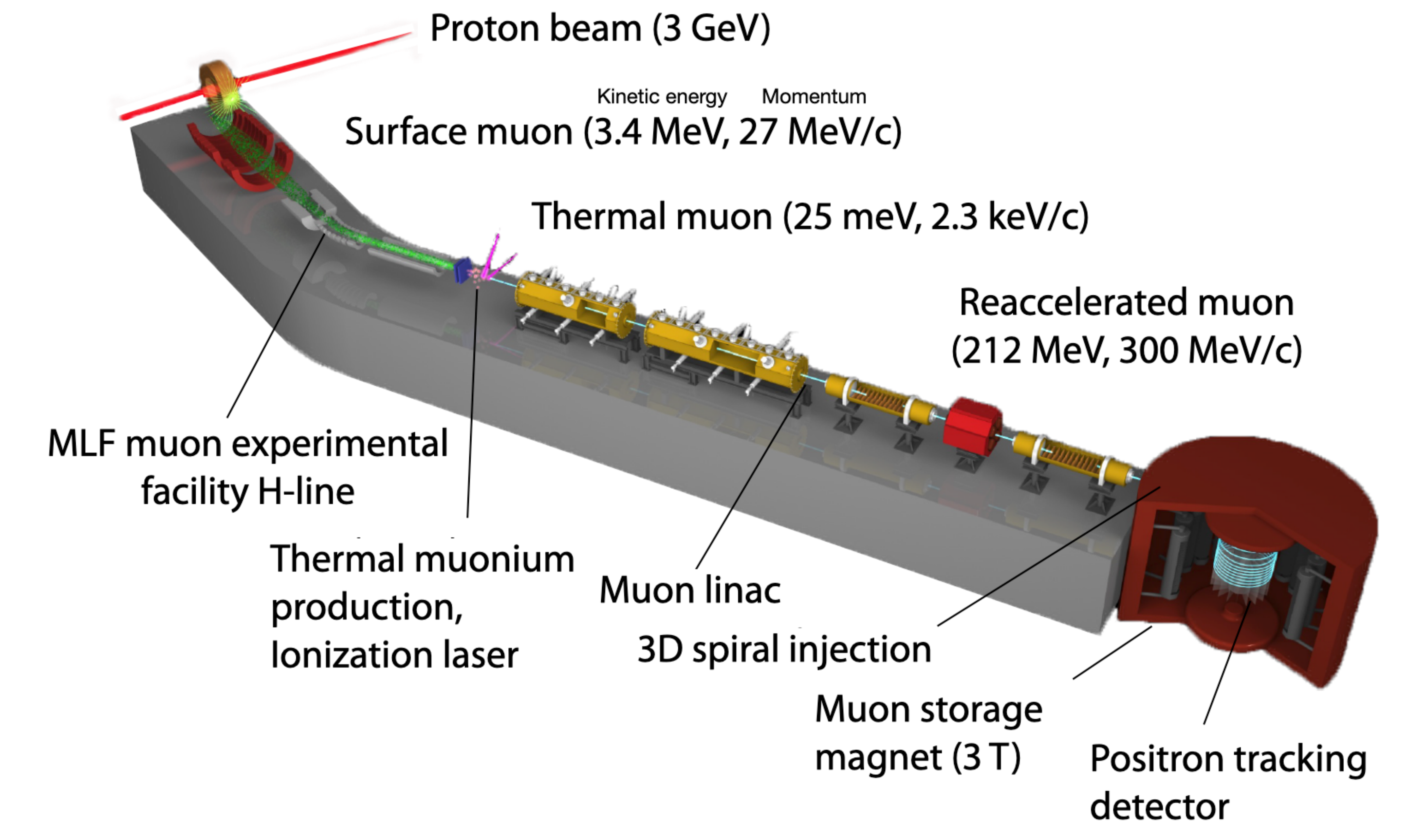}
  \caption{\small An overview of the J-PARC muon $g-2$/EDM experiment at MLF, starting from the 3 GeV proton beam impinging a graphite target to the positron tracking detector in the compact muon storage ring~\cite{Abe:2019thb}.} \label{fig:jparc_overview} 
\end{figure}

The experiment will be conducted at the H-line~\cite{Kawamura:2018apy} in the Materials and Life science experimental Facility (MLF) of J-PARC that is shown in Figure~\ref{fig:jparc_overview} and is currently under construction. A primary 3~GeV proton beam from the J-PARC accelerator will be injected onto a graphite target to produce muon beams. A surface muon beam with a momentum of 29.8~MeV/$c$ will then be extracted to the H-line. The muon beam intensity is expected to be $10^8$ per second with a proton beam power of 1~MW. 

The surface muons will be injected into a room-temperature silica aerogel and will consequently be slowed down, thermalized, and form Mu in the silica aerogel. The thermalized Mu will be diffused to an adjacent vacuum region and ionized by laser irradiation. The emission probability of Mu to the vacuum region is an important factor in increasing statistics. It was found that this probability is enhanced by an order of magnitude by using laser ablation to make holes in the surface of the silica aerogel~\cite{Beer:2014ooa, Beare:2020gzr}. 
The diffused Mu will, at first, be excited from the $1S$ to $2P$ state by irradiating a laser with a wavelength of 122~nm (Lyman-$\alpha$), followed by an irradiation of a laser with a wavelength of 355~nm for the ionization. Another important factor is the ionization efficiency, and this is estimated to be 73\% by assuming the theoretical transition rates and expected laser photon density. 
The original surface muons are nearly 100\% polarized, but the maximum polarization of the produced thermal muons is 50\% as half of the hyperfine states mix $\mu^+$ and $e^-$ spins to give zero muon polarization at ionization.

The resultant thermal muons must be accelerated up to 300~MeV/$c$. They will initially be accelerated up to a kinetic energy of 5.6~keV by a static electric field. A spare radio frequency quadrupole (RFQ) of the J-PARC linear accelerator will then be utilized for subsequent acceleration up to 0.34~MeV in kinetic energy. The muons will finally be further accelerated up to 300~MeV/$c$ by using an interdigital H-type drift tube linear accelerator (IH-DTL), a disk-and-washer structure (DAW) and a disk-loaded traveling wave structure (DLS). Acceleration tests with the RFQ were successfully performed at the D2 area of the D-line in the MLF of J-PARC~\cite{Bae:2018atj}. The IH-DTL and DAW are currently being fabricated.

The low-emittance muons, with thermal energy in the transverse direction and 300~MeV/$c$ in the direction of motion, will then be injected to the muon storage magnet. A 3.0\,T MRI-type superconducting solenoidal magnet will be used for the muon storage~\cite{Sasaki:2016}. The muons will be confined within a cyclotron radius of 33~cm in a 3.0\,T magnetic field. The magnetic field of the muon storage region should be highly uniform as the homogeneity directly affects the experimental sensitivity. The requirement for integrated uniformity along the muon orbit is less than 100~ppb peak-to-peak. Instead of a conventional injection used in the Fermilab experiment, a three-dimensional spiral injection method has been developed due to a space limitation of the J-PARC storage magnet~\cite{Iinuma:2016zfu}. The muons will be injected from the top of the magnet and transported to the storage region by controlling the radial component of the magnetic field. The muons will then be vertically kicked to inside the storage region by two pairs of one-turn coils. The muons will finally be stably stored in the storage region by a weak magnetic focusing field. The injection efficiency is estimated to be 85\%. NMR probes will be used for the magnetic field measurement. Several probes are mounted on the moving stage, and precisely map the magnetic field in the storage region. Additional NMR probes will also be installed below the storage region to measure the time variation of the magnetic field. 

The compact muon storage region enables full tracking of the decay positron using a silicon strip sensor. The positron tracking detector~\cite{Yoshioka:2021bun} will be located in the muon storage magnet separated from the muon storage region by a thin polyimide film. The detector will precisely measure the decay time of the muon by reconstructing the positron track from the muon decay. 

The hit rate will change by approximately two orders of magnitude during the measurement owing to muon lifetime, and the detector will be stable against this. Most of the detector components, including the silicon strip sensor and readout ASIC~\cite{Sato:2020pqt, Kishishita:2020}, are currently in production, and the prototyping of the positron tracking detector has commenced in 2021. 

\begin{figure}[htbp]
  \centering
    \begin{tabular}{c}
      \begin{minipage}{0.5\hsize}
        \includegraphics[width=1.0\linewidth, angle=0]{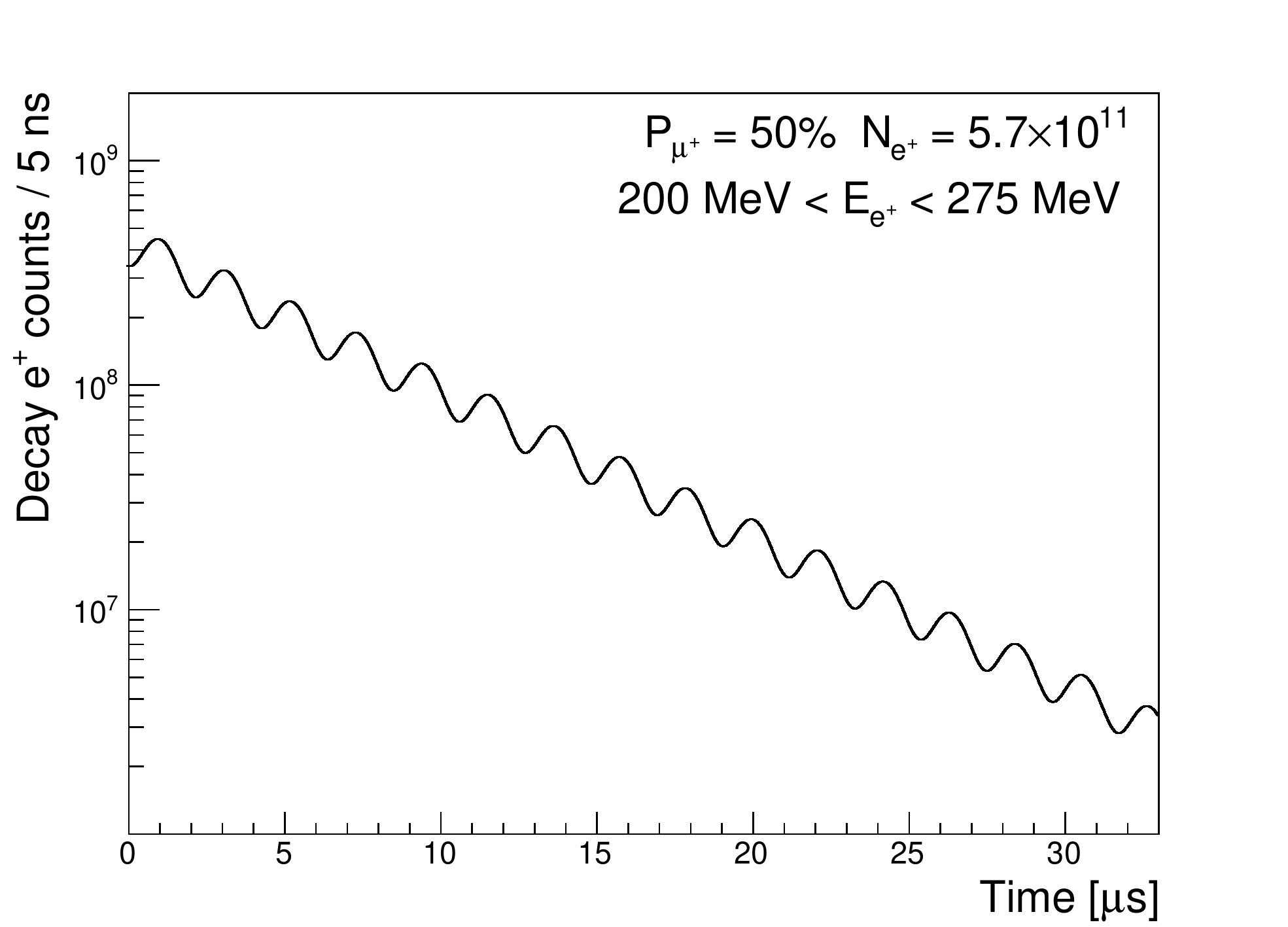}
      \end{minipage}
      \begin{minipage}{0.5\hsize}
        \includegraphics[width=1.0\linewidth, angle=0]{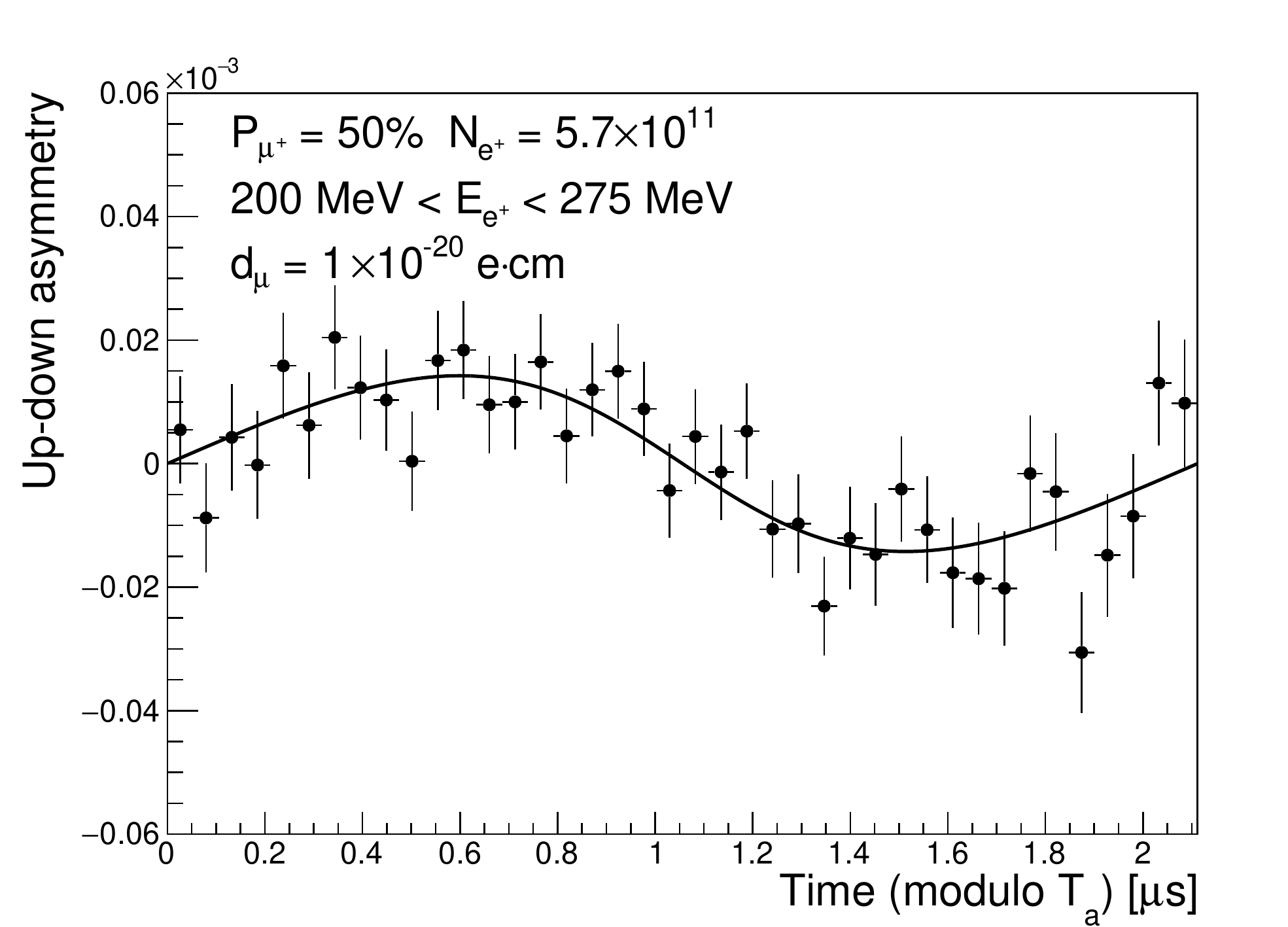}
      \end{minipage}
    \end{tabular}
    \caption{Simulated time distribution of reconstructed positrons (left) 
      and the up-down asymmetry (right). These plots have been reproduced from~\cite{Abe:2019thb}.}
    \label{fig:Wiggle}
\end{figure}

The reconstruction algorithm of the positron track is being developed using Geant4-based simulation data. Using a Hough transformation, a straight line is searched for in the $\phi-z$ plane as a candidate for a high-momentum positron track (where $\phi$ is the angle around the vertical $z$ axis) and a Kalman filter is employed for the track fitting to obtain the track momentum. Using the current algorithm, a track reconstruction efficiency of more than 90\% is found in positron energies ranging between 200~MeV and 275~MeV, even in the expected highest hit rate environment. The number of reconstructed positrons is plotted as a function of the muon decay time as shown in the left figure of Figure~\ref{fig:Wiggle}, and the value of $\omega_a$ is then extracted by fitting to the data. As indicated in the figure, the total number of positrons in the positron energy range between 200~MeV and 275~MeV is expected to be $5.7\times 10^{11}$ for $2.2\times 10^7$ seconds of data taking with a proton beam power of 1~MW. The statistical uncertainty on $\omega_a$ is estimated to be 450~ppb. This directly defines the statistical uncertainty of $a_{\mu}$, because the statistical uncertainty from $\omega_p$ extracted using the NMR probes is negligibly small. The total systematic uncertainty on $a_{\mu}$ is estimated to be less than 70~ppb, which is seven times smaller than the statistical uncertainty.

\subsection{Other experiments}

While no future experiments are currently planned other than the J-PARC Muon $g-2$ experiment, several new ideas based on at ``magic momentum", higher than ``magic momentum" and lower than ``magic momentum" approaches have been discussed in recent years. A summary of such proposed experiments is given in Table~\ref{tab:comparison}. To further improve the precision of the ``magic momentum" approach, a more intense muon beam coupled with better control of systematic effects arising from the precession frequency measurement, magnetic field measurement, and spatial and temporal motion of the beam are essential. The Heavy Ion Accelerator Facility (HIAF)~\cite{Yang:2013yeb} in Huizhou, China has performed a study for an intense muon source~\cite{Cai:IPAC2019-THPGW041} using a heavy-ion on-target approach that is very similar to the one at Fermilab (which uses protons on target). Based on a study using $^{136}$Xe-on-target at HIAF-U, the expected muon intensity is roughly 30 times higher than that of Fermilab's beamline for the Muon $g-2$ experiment~\cite{Sun:2020}. A factor of two improvement in the total uncertainty could be achieved with improvements in the beam storage, positron detection and beam detection systems.

\begin{table}[htbp]
\begin{center}
\begin{tabular}{cccccc} 
\toprule 
 Experiment & Ref. & Storage conditions & Beam  \\ \hline 
 BNL/Fermilab & \cite{Grange:2015fou} & $R=7.1$\,m storage ring, $B=1.45$~T & 3.1\,GeV/c  \\  
 J-PARC & \cite{Abe:2019thb} & $R=0.35$\,m diameter storage ring, $B=3.0$~T & 0.3\,GeV/c    \\  
 Farley03 & \cite{Farley:2003mj} & Noncontinuous uniform magnets, $B=1.47$~T & 15\,GeV/c    \\ 
 Silenko10 & \cite{Silenko:2010fe} & Noncontinuous nonuniform magnets, $B=$ arbitrary & arbitrary    \\  
 PSI (muEDM) & \cite{himb:2021} & Solenoid, $B=6$~T & 0.125\,GeV/c  \\ \bottomrule
\end{tabular}
\caption{Current and future proposals for experiments to measure the muon $g-2$.}
\label{tab:comparison}
\end{center}
\end{table}

In~\cite{Farley:2003mj}, a proposal for a novel method could potentially reach 30\,ppb using a set of magnets (non-continuous) with uniform field and edge focusing. The main distinctions between this approach and the BNL/FNAL approaches are: i) the measurement of the average magnetic field is achieved using a polarized proton beam instead of protons at rest in the NMR probes, and (ii) a muon momentum much higher the ``magic momentum", i.e., a 15~GeV muon beam. The prolonged lab frame lifetime of the muon increases the measurement time and thus the precision. A similar idea was also contemplated in~\cite{Silenko:2010fe} utilizing a usual storage ring with a non-continuous, non-uniform magnetic field, and magnetic focusing. A facility that could produce a secondary muon beam at 15~GeV is the CERN PS East Area~\cite{Durieu:1997kc}. The main proton driver is 24~GeV/c and the secondaries are from 0.5~GeV/c to 15~GeV/c. The M2 muon beamline at the SPS is another facility at CERN that could produce very high energy muon beam~\cite{Doble:1994np,Sba:2021}. The beam is produced by protons impinging on a primary beryllium target and transported to the experimental area in an evacuated beam line. Currently, the beamline is designed to transport high fluxes of muons with momenta in the range between 100~GeV/c and 225~GeV/c that could be derived from a primary proton beam of 400~GeV/c with intensity between $10^{12}$ and $10^{13}$ protons per SPS spill. A new approach in creating a compact muon beam is with the laser Wakefield acceleration of electrons to a few GeV and then colliding them with a target~\cite{Xu:2020, Yu:2018omk,Gonsalves:2019wnc,Boscolo:2018ytm}. Preliminary results are promising for this approach. The produced low-energy muon beam can be further accelerated using another Wakefield acceleration. In a study performed in~\cite{Wang:2021hfr}, a 275\,MeV muon beam can be accelerated to more than 10\,GeV within 22.5\,ps.

The high-intensity muon beamline (HiMB) project currently under development at PSI aims to provide a 100 times higher surface muon rate of $10^{10}~\mu^{+}$/s at 28\,MeV/c by upgrading production targets and beamlines~\cite{Berg:2015wna}. Combined with the longitudinal and transverse muon cooling technique developed at the same institute will produce a low energy polarized muon beam with low emittance~\cite{Bao:2014xxa,Antognini:2018bgr,Iwai:2020jye,Antognini:2020uyp}. The muEDM collaboration at PSI proposed to utilize the HiMB and the same apparatus for their proposed muon EDM search (see Section~\ref{sec:muEDM}) to measure the muon $g-2$~\cite{himb:2021}. By configuring the solenoid to 6~T and collecting data for a year, it is possible to reach a statistical sensitivity of 0.1~ppm.

\section{Related experiments}\label{sec:relatedexps}

\begin{figure}[htbp]
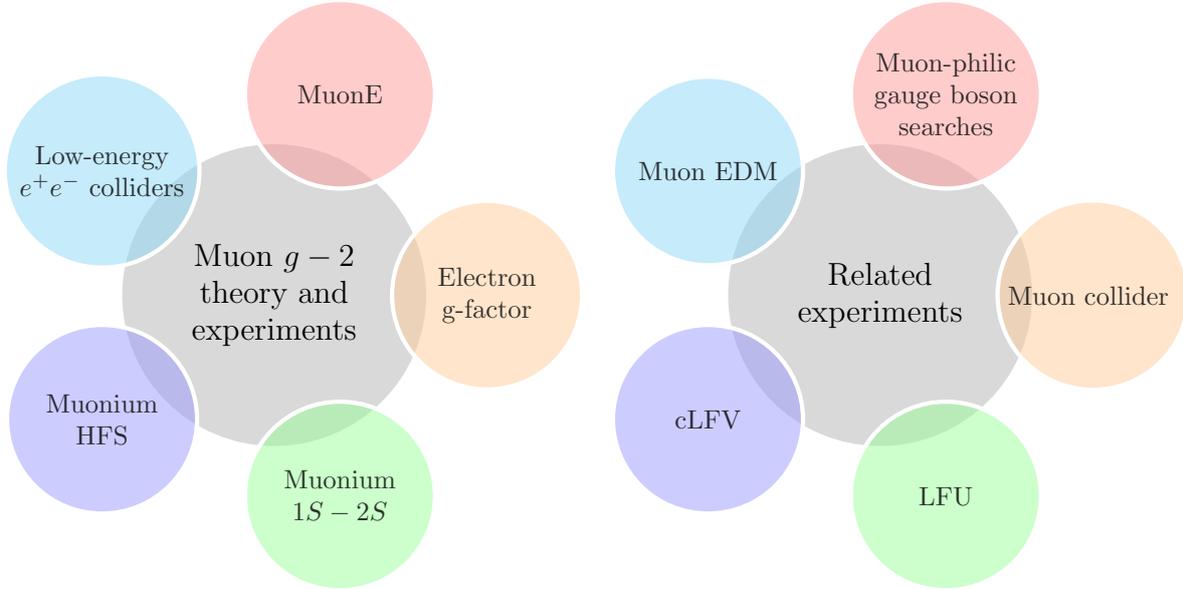

\centering
\smartdiagram[bubble diagram]{
Muon $g-2$~\\theory and\\experiments, MuonE, Low-energy\\$e^{+}e^{-}$ colliders, Muonium\\HFS, Muonium\\$1S-2S$, Electron\\g-factor
}
\smartdiagram[bubble diagram]{
Related~\\experiments,Muon-philic\\gauge boson\\searches, Muon EDM, cLFV, LFU, Muon collider
}
\caption{\small An overview of experiments directly and indirectly connected to the physics of muon $g-2$.}
\label{fig:MuonG2Physics-Diagram}
\end{figure}

This section provides a comprehensive overview of planned and ongoing experiments that will help improve the precision of both the theoretical predictions and experimental measurements of $a_\mu$. Also discussed are related experiments that will provide complementary information regarding BSM physics related to the muon $g-2$. A pictorial representation of the related experiments mentioned here is shown in Figure~\ref{fig:MuonG2Physics-Diagram}.

\subsection{MuSEUM (J-PARC)}

A Muonium (Mu) ground-state hyperfine structure (HFS) measurement is closely related to the determination of the muon $g-2$. One of the ratios of fundamental constants in equation~\eqref{eq:amu}, the muon-to-electron mass ratio, $m_{\mu}/m_{e}$ can be extracted from a precise measurement of Mu HFS. The most precise result to-date is $\nu_{\rm{HFS}}=4463.302765(53)$~MHz~\cite{Liu:1999iz} (converted to 120~ppb for $m_{\mu}/m_{e}$). The MuSEUM collaboration at J-PARC plans to improve further the precision with a pulsed muon beam technique~\cite{Shimomura:2011zz}. 

In the MuSEUM experiment, a muon beam polarized anti-parallel to the beam direction is injected into Kr gas to form Mu. The Mu is subsequently irradiated by microwaves and the number of decay positrons from Mu are counted by a positron detector located downstream of the microwave cavity. As a Phase-I experiment, $\nu_{\rm{HFS}}$ is directly measured in a near-zero magnetic field (ZF) condition with three layers of the permalloy magnetic shielding. This ZF experiment is conducted at the D2 area of the D-line in the MLF of J-PARC, where the provided muon beam intensity is $6\times10^6$ per second with 0.6~MW operation. The ZF experiment commenced in 2014 and the first result was found to be $\nu_{\rm{HFS}}=4.463302(4)$~Hz~\cite{Kanda:2020mmc}. With the ZF condition, some data was collected using a silicon strip detector for positron detection (originally developed for the E34 experiment as a prototype~\cite{Aoyagi:2019ujx}), and a new method was developed to directly analyze the Rabi oscillation using the data. The final ZF experiment will be conducted in 2021. In the next step, the Zeeman sub-levels in a high magnetic field (HF) will be measured to extract $\nu_{\rm{HFS}}$. The HF experiment will be conducted at the H1 area of the H-line in the MLF and will provide a muon beam intensity of $10^8$ per second with 1~MW beam power. Various R\&D for the HF experiment started in 2011. One of the key elements is a magnet with high uniformity and stability. An MRI-type superconducting magnet that provides a 1.7~T magnetic field will be employed for the experiment. Passive shimming was performed for the magnet, and 0.17~ppm peak-to-peak uniformity has been achieved, which satisfies the requirement. NMR probes to precisely measure the magnetic field have been developed in close collaboration with the E34 experiment. The development of an additional positron counter to be installed upstream of the microwave cavity is also ongoing, with the purpose of reducing the systematics related to the muon stopping distribution. The HF experiment is planning to start after the H-line construction is finished and aims to reach 5~Hz statistical precision with 40~days of measurement.

\subsection{Muonium 1S-2S (Mu-MASS)}

The Mu-MASS experiment~\cite{Crivelli:2018vfe,Ohayon:2021dec} aims to measure the $1S$-$2S$ transition frequency of muonium ($\Delta \nu_{1S2S}$) to an unprecedented precision of 10~kHz (4 ppt level), a 1000-fold improvement over previous measurements~\cite{Fan:2014kza}. The experiment relies on newly-developed cryogenic 100~K muonium sources, new excitation and detection schemes implemented for positronium spectroscopy, and advancements in UV lasers. It is planned to be performed at PSI using the low-energy muon beamline (LEM) at $\mu$E4~\cite{Prokscha:2008zz} and will provide the best determination of the muon-to-electron mass ratio at 1~ppb, critical for future muon $g-2$ experiments. Several milestones were recently achieved for the experiment~\cite{Crivelli:2021} including the demonstration of $2S$-state detection, detection of a bound electron from Mu in coincidence with the positron, a circulating power of 20\,W for the laser system (improvement in stability is ongoing) and establishing a suitable reference frequency for the experiment. Phase-1 of the experiment could reach a 100~kHz uncertainty after 10 days of data taking. In Phase-2, an uncertainty of 10~kHz could be reached after 40 days of data taking.

\subsection{Muonium 1S-2S (J-PARC)}

A measurement of $\Delta \nu_{1S2S}$ is also currently being prepared at J-PARC~\cite{Uetake:2019jparc, Zhang:2021cba}, with the aim of achieving a precision more than two orders of magnitude better than previous measurements~\cite{Fan:2014kza}.  The experiment will be conducted at the S2 area of the S-line in the MLF of J-PARC, which will be in operation in 2021. The silica aerogel developed for the E34 experiment is used for the Mu production target. The surface muon beam produced at the muon production target in the MLF is injected to the silica aerogel to form Mu, and then the thermalized Mu is emitted from the aerogel. An excitation from $1S$ state to $2S$ state is made by irradiating a continuous-wave (CW) laser with a wavelength of 244~nm to the Mu followed by an electron dissociation by a pulsed YAG laser with a wavelength of 355~nm. The resultant thermal muon is accelerated by a static electric field and detected by a micro channel plate detector (MCP) located at the most downstream end of the transporting line. The MCP consists of an electrostatic mirror and bending magnet to remove backgrounds (e.g., positrons) in the surface muon beam. The beam transport efficiency has been evaluated to be 60\% using Geant4 simulations. An external cavity diode laser with a wavelength of 976~nm is used as a master oscillator, and a 244~nm beam is produced after several amplifiers and wavelength conversions. A pulsed laser system is also currently being developed for quick confirmation of the $1S$-$2S$ transition as a Phase-0 experiment aiming to be conducted in early 2022. The pulsed laser system can also be applied as an alternative for the thermal muon production of the E34 experiment instead of the current primary option of the Lyman $\alpha$ laser. The Phase-1 experiment, aiming to achieve 1~MHz precision with the CW laser system, is planned to start in late-2022. As part of a Phase-2 experiment, the development of a high-power CW laser will facilitate reaching 100~kHz in precision within a couple of years.

\subsection{Searches for muon-philic gauge bosons}\label{sec:darkZ}

The current 4.2$\sigma$ discrepancy in muon $g-2$ can be explained by the existence of a new dark boson $Z_{\mu}$ with a mass in the sub-GeV range and which couples predominantly to the second and third lepton generations through the $L_{\mu}-L_{\tau}$ current~\cite{Gninenko:2014pea}. This $Z_{\mu}$ can, in principle, be produced via muon scattering off nuclei and the signal can be searched for via $Z_{\mu}\rightarrow \nu\nu$ process. An upgraded version of the NA64 experiment at CERN, NA64$\mu$, is capable of searching for such a signature via a missing muon beam energy approach~\cite{Gninenko:2014pea}. The main goal of the experiment in the 2021 pilot run with the 100 GeV muon beam at M2, SPS of CERN was to commission the NA64$\mu$ detector and to explore the still unexplored area of the coupling strengths and masses of the dark boson $Z_{\mu}$ that could still explain the muon $g-2$ anomaly~\cite{Gninenko:2020hbd,Sieber:2021fue}. A similar experiment, M$^{3}$ (Muon Missing Momentum) is also being proposed at Fermilab~\cite{Kahn:2018cqs}. The experiment aims to achieve $10^{10}$ muons on targets in Phase-I and $10^{13}$ muons on targets in Phase-II.

\subsection{Muon electric dipole moment (EDM)}\label{sec:muEDM}

In general, the anomalous magnetic moment $a_{\mu}$ and electric dipole moment $d_{\mu}$ can be considered as the real and imaginary parts, respectively, of a general dipole operator~\cite{Feng:2001sq} which can be connected through certain parameterizations~\cite{Feng:2001sq}. The current limit on the muon EDM of $|d| < 1.9\times10^{-19}~e\cdot\rm{cm}$ (95\% C.L.) is provided by the BNL muon $g-2$ collaboration~\cite{Bennett:2008dy}. 
For $\vec{\beta}\cdot\vec{B}=0$ and the case of ``magic momentum" or ``no focusing electric field", equation~\eqref{eq:omega_aFull} can be written as
\begin{equation}\label{eq:omegaDiff}
    \vec{\omega}_{\rm{diff}} = -\frac{q}{m_{\mu}}\Bigg[ a_{\mu} \vec{B} + \frac{\eta}{2}\left(\frac{\vec{E}}{c}+\vec{\beta}\times\vec{B}\right)\Bigg] = \vec{\omega}_{a} + \vec{\omega}_{\rm{EDM}}
\end{equation}
where $\eta$ is a dimensionless parameter directly proportional to the size of the muon EDM. In a typical storage ring experiment where the dipole magnetic field is in the vertical direction and the muon is orbiting in the horizontal plane, the vector $\vec{\omega}_{a}$ and $\vec{\omega}_{\rm{EDM}}$ are orthogonal to each other. As a result, the precession plane of the muon spin will be tilted by an angle $\delta=\tan^{-1}(\frac{\eta\beta}{2a_{\mu}})$. As the decay positron is emitted preferentially in the direction of the muon spin, the vertical positron angle of the decay will oscillate as a function of time.

The Fermilab experiment will search for the muon electric dipole moment (EDM)~\cite{Chislett:2016jau} by measuring the tilt of the $g-2$ precession plane. Both tracking-based and calorimetry-based measurements will be performed for the EDM search. While the tracking-based analysis will look for an oscillation in the vertical positron angle, the calorimetry-based analysis will look for i) an oscillation in the mean vertical hit position of the positron out of phase with the $g-2$ oscillation, and ii) a distortion in the ``V-shape" of the $g-2$ phase versus vertical hit position. These new measurements will provide complementary information about possible BSM physics in the muon sector that may be directly associated with the muon magnetic anomaly. The expected sensitivity is of the order of $10^{-21}e$~cm.

Similarly, in the J-PARC muon g-2/EDM experiment, a muon EDM can be simultaneously probed by extracting the up-down asymmetry of the number of reconstructed positrons. The value of the muon EDM of $1\times10^{-20}~e\cdot$cm is assumed in the right figure of Figure~\ref{fig:Wiggle}, and the statistical sensitivity from the simulated data is estimated to be $1.5\times 10^{-21}~e\cdot$cm~\cite{Abe:2019thb}. The systematic uncertainty on the EDM is estimated to be four times smaller than the statistical uncertainty, mainly due to the misalignment of the positron tracking detector.

The muEDM collaboration~\cite{Adelmann:2021udj,Sakurai:2022tbk,Khaw:2022igy} at PSI aims to perform a sensitive search of the muon electric dipole moment at the $10^{-23}$~e$\cdot$cm level using the frozen-spin technique. In this approach, a radial electric field $E_{r}\sim aBc\beta\gamma^2$ is chosen such that only the $\vec{\omega}_{\rm{EDM}}$ term remains in equation~\eqref{eq:omegaDiff}. As a result, there is no anomalous precession in the storage plane and the vertical polarization of the muon will increase over time. To search for an EDM signal, the up-down asymmetry of the positron count is measured using upper and lower detectors. The improved result would provide valuable complementary insights and support the search for BSM physics in lepton magnetic moments~\cite{Crivellin:2018qmi,Crivellin:2019mvj}. If the current muon $g-2$ tensions were to persist, a measurement of the muon EDM would help disentangle the flavor structure of an underlying BSM scenario.

\subsection{Searches for charged lepton flavor violation (cLFV)}\label{sec:cLFV}

As one of the key observables in particle physics, lepton flavor violation in muon decays ($\mu \rightarrow e\gamma$, $\mu \rightarrow eee$), nuclear $\mu-e$ transitions and muonium to antimuonium conversion are closely connected to the anomalous magnetic moment of the muon. Various BSM models such as MSSM, $B-L$ and $L_{\mu}-L_{\tau}$ are studied extensively in \cite{Lindner:2016bgg}. While limits from collider experiments, MEG and nuclear $\mu-e$ transitions have severely constrained parameter spaces that could accommodate both the muon $g-2$ and cLFV, there remain regions where BSM physics are allowed. Future experiments that could further constrain the BSM phase space are summarized below:
\begin{itemize}
    \item $\mu \rightarrow e\gamma$: the MEG collaboration at PSI published a final result in 2016 with a limit of $B(\mu \rightarrow e\gamma) < 4.2 \times 10^{-13}$ (90\% C.L.). An upgraded version of the experiment with a sensitivity of $6 \times 10^{-14}$, MEG~II~\cite{Baldini:2018nnn}, has started its engineering run during 2020-2021 and will be followed by three years of physics data taking~\cite{Chiappini:2020ufo}. 
    \item $\mu \rightarrow eee$: The Mu3e collaboration~\cite{Blondel:2013ia} at PSI was formed in 2013 with an aim of reaching the sensitivity to a branching ratio of one in $2\times10^{15}$ in Phase-I and one in $10^{16}$ muon decays for Phase-II. The experiment is currently under construction and will take place at the Paul Scherrer Institute (PSI) in Switzerland. In order to achieve the targeted number of muon decays, the collaboration will utilize the very intense $10^{8}$~$\mu^{+}$/s $\pi$E5 beamline (Phase-I) and a new high-intensity muon beam line (HiMB) that will provide $10^{9}-10^{10}$~$\mu^{+}$/s (Phase-II).
    \item Nuclear $\mu-e$ transitions: The most stringent limit on these transitions is set by the SINDRUM II collaboration~\cite{Bertl:2006up} for $\mu \rightarrow e$ in gold: $R_{\mu e} < 7 \times 10^{-13}$ (90\% confidence level). There are currently two planned experiments searching for $\mu^{-} N \rightarrow e^{-}N$ transitions using different nuclei; the Mu2e collaboration at Fermilab~\cite{Carey:2008zz, Bartoszek:2014mya} and the COMET collaboration at J-PARC~\cite{Cui:2009zz, Kuno:2013mha}. Mu2e is projected to reach a maximum sensitivity of $7 \times 10^{-17}$, whilst COMET Phase-I will reach a sensitivity of $7\times10^{-15}$~\cite{COMET:2018auw}. Both Mu2e and COMET are actively pursuing upgrades~\cite{Mu2e:2018osu,COMET:2018auw} to their experiments to increase the sensitivity by an order and two orders of magnitude, respectively.
    \item Muonium to antimuonium conversion: The best upper limit of $P(\rm{Mu}\leftrightarrow\overline{\rm{Mu}})<8.3 \times 10^{-11}$ (90\% confidence level) is set by the PSI experiment in 1999~\cite{Willmann:1998gd}. The MACE collaboration~\cite{Tang:2021} at the Chinese Spallation Neutron Source (CSNS) is aiming for a $10^{-14}$ sensitivity using the high quality intense slow muon sources with more than $10^{8}~\mu^{+}$ produced per second and the beam spread smaller than 5\% at CSNS~\cite{Han:2021nod}.
\end{itemize}

\subsection{Tests of lepton flavor universality (LFU)}

Recent hints for the violation of CKM unitarity, together with the anomalies exhibited by semi-leptonic B-meson decays, B-meson decays of $B^{+}\rightarrow K^{+}l^{+}l^{-}$ and the muon and electron $g-2$, can be viewed as further evidence for lepton flavor universality violation (LFUV). The LHCb experiment recently reported a new measurement of the double ratio of branching fractions~\cite{Aaij:2021vac}:
\begin{equation}
    R_K = \frac{\mathcal{B}(B^+\rightarrow K^+\mu^+\mu^-)}{\mathcal{B}(B^+\rightarrow J/\psi (\rightarrow \mu^+\mu^-) K^+)} \bigg{/} \frac{\mathcal{B}(B^+\rightarrow K^+e^+e^-)}{\mathcal{B}(B^+\rightarrow J/\psi (\rightarrow e^+e^-) K^+)} = 0.846 ^{+0.044}_{-0.041} \, ,
\end{equation}
leading to a $3.1\sigma$ deviation from the SM prediction of unity as evidence of a potential violation of lepton flavor universality. 

Future experiments such as the PIONEER~\cite{Mazza:2021adt} can also provide very sensitive tests of lepton flavor universality~\cite{Crivellin:2020lzu}. The PIONEER experiment aims to measure the ratio
\begin{equation}
    R_{e/\mu}=\frac{\Gamma(\pi^{+}\rightarrow e^{+}\nu(\gamma))}{\Gamma(\pi^{+}\rightarrow \mu^{+}\nu(\gamma))}
\end{equation}
of pion decays to test the $e-\mu$ universality in charged-current weak interactions. The SM prediction has reached $10^{-4}$ level precision at $R_{e/\mu}=1.2552(2) \times 10^{-4}$~\cite{Cirigliano:2007xi,Cirigliano:2007ga, Bryman:2011zz}, whereas the experimental value from the PiENu Collaboration is less precise, at $R_{e/\mu}=1.2344(30) \times 10^{-4}$~\cite{Aguilar-Arevalo:2015cdf}. A potential location for the experiment is PSI, where the high intensity pion beamlines of $\pi$E1 and $\pi$E5 are available. The key feature for the PINOEER experiment is a highly segmented active target coupled with Low Gain Avalanche Detectors (LGAD) technology~\cite{Pellegrini:2014lki}. Two detector options currently under consideration are a liquid-xenon scintillation calorimeter and a LYSO scintillation calorimeter~\cite{Mazza:2021adt}.

\subsection{Muon collider}

Many scenarios are being considered to physics related to the discrepancy of muon $g-2$ using a multi-TeV muon collider. While the technology to perform such measurements are currently out of reach, possible new physics scenarios can be directly explored using TeV-scale muon colliders such as the gauged $L_{\mu}-L_{\tau}$ model~\cite{Huang:2021nkl}, BSM singlets~\cite{Capdevilla:2020qel, Capdevilla:2021rwo}, dimension-6 operator in the Standard Model Effective Field Theory (SMEFT)~\cite{Buttazzo:2020ibd, Cheung:2021iev} and muon-smuon-bino like system in minimal supersymmetric SM (MSSM)~\cite{Yin:2020afe}.

\section{Summary and conclusions}

Confirmation of the BNL measurement of the muon $g-2$ by the Fermilab Muon $g-2$ collaboration has strengthened the evidence for BSM physics. Together with other muon experiments searching for lepton flavor violation, an electric dipole moment, tests of lepton universality, and flavor anomalies in meson decays, the possibilities for further probing new physics scenarios are extensive. Table~\ref{tab:schedule_summary} summarizes the physics goals and best known schedules of experiments mentioned in the article.

\begin{table}[htbp]
\centering
\small
\begin{tabular}{ccccc} \toprule 
 Experiment & Current Status & Physics Run & Physics Goal & Reference \\ \hline 
 Muon $g-2$ (Fermilab) & Data taking & 2018 & $\delta a_{\mu}/a_{\mu}=140$~ppb & \cite{Grange:2015fou}\\
 Muon $g-2$/EDM (J-PARC) & Prototyping & 2027 &$\delta a_{\mu}/a_{\mu}=450$~ppb & \cite{Abe:2019thb}\\
 MuSEUM (J-PARC) & Phase-I completed & 2014 & $\delta \nu_{HFS}/\nu_{HFS} = 1~\rm{ppb}$ & \cite{Shimomura:2011zz,Kanda:2020mmc}\\
 Mu $1S-2S$ (PSI)& Engineering Run & 2022 & $\delta \nu_{1S-2S}/\nu_{1S-2S} = 4~\rm{ppt}$& \cite{Crivelli:2018vfe,Crivelli:2021,Ohayon:2021dec} \\
 Mu $1S-2S$ (J-PARC)& Engineering Run & 2022 & $\delta \nu_{1S-2S}/\nu_{1S-2S} = 40~\rm{ppt}$& \cite{Uetake:2019jparc,Zhang:2021cba}\\
 MEG II (PSI) & Engineering Run & 2022 & $Br(\mu \rightarrow e\gamma)\sim6\times10^{-14}$& \cite{Baldini:2018nnn,Chiappini:2020ufo}\\
 Mu2e (Fermilab) & Construction & 2025 & $R_{\mu e}\sim 7 \times 10^{-17}$ & \cite{Bartoszek:2014mya} \\
 COMET phase-I (J-PARC) & Construction & 2023 & $R_{\mu e}\sim 7 \times 10^{-15}$ & \cite{COMET:2018auw}\\
 Mu3e phase-I (PSI) & Construction & 2025 & $Br(\mu^{+} \rightarrow e^{+}e^{+}e^{-}) \sim 2 \times 10^{-15}$ & \cite{Blondel:2013ia}\\ \hline
 MUonE (CERN) & R\&D & - &$\delta a^{HVP}_{\mu}/a^{HVP}_{\mu}=0.5\% $ & \cite{Calame:2015fva,Abbiendi:2016xup,Abbiendi:2677471}\\
 NA64$\mu$ (CERN) & R\&D & - & $g_{S},g_{V} \sim 10^{-5}$ & \cite{Gninenko:2014pea} \\
 M$^{3}$ (Fermilab) & R\&D & - & $g_{S},g_{V} \sim 10^{-5}$ & \cite{Kahn:2018cqs} \\
 muEDM (PSI) & R\&D & - & $d_{\mu} \sim 6 \times 10^{-23}$ & \cite{Adelmann:2021udj} \\
 MACE (CSNS) & R\&D & - & $P(M\leftrightarrow\bar{M}) \sim 10^{-14}$ & \cite{Tang:2021}\\
 PIONEER (PSI) & R\&D & - &  $R_{e/\mu} \sim 10^{-4}$ & \cite{Crivellin:2020lzu,Mazza:2021adt}\\  \bottomrule
\end{tabular}
\caption{An overview of the goals and expected schedules for experiments mentioned in this article. Experiments where the status is labeled R\&D are currently undergoing feasibility studies while the rest are approved experiments.}
\label{tab:schedule_summary}
\end{table}

\section*{Acknowledgments}
We sincerely thank our colleagues in the Fermilab Muon $g-2$ collaboration, the J-PARC muon $g-2$/EDM collaboration and the Muon $g-2$ Theory Initiative for their useful feedback regarding the manuscript. We would additionally like to thank David Hertzog, Martin Hoferichter, Mark Lancaster, Tsutomu Mibe, Adam Schreckenberger and Dominik St\"{o}ckinger for providing numerous useful discussions and feedback. 

A.K. is supported by STFC under the consolidated grant ST/S000925/1. K.S.K. is supported by the National Natural Science Foundation of China (Grant No. 12075151 and 12050410233). T.Y. is supported by the Japan Society for the Promotion of Science (JSPS) KAKENHI Grant Nos. JP20H05625 and JP15H05742.the the the the 

\bibliography{mybibfile}
\biboptions{sort&compress}

\end{document}